\documentclass[times,final]{elsarticle}
\pdfoutput=1

\usepackage{subcaption}

\usepackage{amsmath}
\usepackage{amssymb}

\usepackage{tikz}
\usepackage{quiver}
\usepackage{adjustbox} 

\usepackage{cancel}

\usepackage{siunitx}

\usepackage{physics}

\usepackage{mathtools}

\DeclareMathOperator{\AD}{AD}
\DeclareMathOperator{\Ad}{Ad}
\DeclareMathOperator{\ad}{ad}

\newcommand{\bE}{\mathbb{E}}

\newcommand{\bI}{\mathbb{I}}

\newcommand{\bR}{\mathbb{R}}

\newcommand{\cB}{\mathcal{B}}
\newcommand{\cC}{\mathcal{C}}
\newcommand{\cD}{\mathcal{D}}
\newcommand{\cE}{\mathcal{E}}

\newcommand{\cO}{\mathcal{O}}

\newcommand{\cQ}{\mathcal{Q}}

\newcommand{\cX}{\mathcal{X}}

\newcommand{\gset}[1]{\mathbf{#1}}

\newcommand{\QuantitySet}{\gset{Q}}

\usepackage[unicode=true]{hyperref}

\hypersetup{%
	plainpages=false,
	bookmarksopen=true,
	bookmarksnumbered=true,
	breaklinks=true,
}

\usepackage{amsthm}

\usepackage[capitalize,nameinlink,noabbrev]{cleveref}

\newtheoremstyle{break} 
  {}          
  {}          
  {}          
  {}          
  {\bfseries} 
  {.}         
  {\newline}  
  {}          
\theoremstyle{break}

\newtheorem*{remarkxx}{Remark}
\newenvironment{remark*}
  {\pushQED{\qed}\remarkxx}
  {\popQED\endremarkxx}

\usepackage{booktabs} 

\graphicspath{{figures/}}

\journal{arXiv}

\newcommand{\titel}{%
  Exergetic Port-Hamiltonian Systems
  \texorpdfstring{\\}{}
  for Multibody Dynamics
}

\hypersetup{%
  pdfauthor={%
    Markus Lohmayer
    Giuseppe Capobianco
    Sigrid Leyendecker
  },
	pdftitle=\titel
}

\begin{document}

\begin{frontmatter}
	\title{\titel}

	\author[fau]{Markus Lohmayer}
	\ead{markus.lohmayer@fau.de}

	\author[fau]{Giuseppe Capobianco}

	\author[fau]{Sigrid Leyendecker}

	\address[fau]{%
		Institute of Applied Dynamics\\
   	Friedrich-Alexander Universität Erlangen-Nürnberg,
		Erlangen, Germany
	}

	\begin{abstract}
%
Multibody dynamics simulation plays an important role
in various fields, including
mechanical engineering, robotics, and biomechanics.
%
%
Setting up computational models
however becomes increasingly challenging
as systems grow in size and complexity.
Especially
the consistent combination of models
across different physical domains
usually demands a lot of attention.
%
%
This motivates us
to study
formal languages for
compositional modeling of
multiphysical systems.
%
%
This article shows how
multibody systems,
or more precisely
assemblies of rigid bodies connected by lower kinematic pairs,
fit into the framework of
Exergetic Port-Hamiltonian Systems (EPHS).
%
%
This approach is based on
the hierarchical decomposition of systems
into their ultimately primitive components,
using a simple graphical syntax.
Thereby, cognitive load can be reduced
and communication is facilitated, even with non-experts.
Moreover,
the encapsulation and reuse of subsystems
promotes efficient model development and management.
%
%
In contrast to established modeling languages such as Modelica,
the primitive components of EPHS are not defined by arbitrary equations.
Instead,
there are four kinds of components,
each defined by a particular geometric structure
with a clear physical interpretation.
This higher-level approach could make the process of
building and maintaining large-scale models simpler and also safer.

	\end{abstract}

\end{frontmatter}

\section{Introduction}%

\subsection{EPHS modeling language}%

Exergetic Port-Hamiltonian Systems (EPHS),
as recently formalized
in~\cite{2024LohmayerLynchLeyendecker},
provide
a compositional modeling language
for physical systems.
Specifically,
the language is designed for
the efficient combination of
dynamic models from
classical mechanics,
electromagnetism,
and
irreversible processes
(with local thermodynamic equilibrium).

The central paradigm is that
systems are in general \emph{systems of systems}.
In other words,
a model of a physical system
is often best understood as
an interconnection of simpler models.
To make this intuitive,
a graphical syntax is used to
define interconnections.
An expression in the EPHS syntax
is hence called an \emph{interconnection pattern}.
\Cref{fig:mbs} shows an example.
The round \emph{inner boxes}
of such formal diagrams
correspond to
(interfaces of) subsystems,
while the rectangular \emph{outer box} represents
the interface of the composite system.
The black dots are called \emph{junctions}
and the connected lines are called \emph{ports}.
An important feature of the language is that
each junction corresponds to an energy domain,
such as
the potential or the kinetic energy domain
(of a rigid body).
Since ports expose energy domains,
the mantra of EPHS is that
systems are interconnected by
sharing energy domains.
This is reflected
on the level of a single interconnection pattern
and also by the fact that
there is a unique way to
flatten any hierarchy of nested interconnection patterns.
Two interconnection patterns compose,
i.e.~they can be nested,
whenever the outer interface of one pattern
represents the same interface as
some inner box of the other pattern.
The syntax hence provides
a graphical abstraction for
dealing with
interconnected and hierarchically nested systems
in terms of their interfaces.

Any system is either
a primitive system
or a composite system
(whose subsystems can be
primitive or composite systems).
There are different kinds of primitive systems
representing either
energy storage,
reversible energy exchange or
irreversible energy exchange.
The semantics of a primitive system
is accordingly determined by a geometric object,
whose structure reflects
the fundamental thermodynamic behavior.
The semantics of a composite system
is determined by
the semantics of its subsystems
and their interconnection.
The interconnection is given by
the respective interconnection pattern
with each junction implying
shared state and a power balance.
Due to the structural properties of
primitive systems and interconnections,
all systems conform with
the first and second law of thermodynamics,
Onsager reciprocal relations,
and some further conservation laws (e.g.~for mass).
Besides ensuring thermodynamic consistency,
restricting the set of possible models in this way
is a key enabling factor for
their easy reusability.

\subsection{Related work}%

The relationship between
the EPHS modeling language,
bond graph modeling,
port-Hamiltonian systems,
and the GENERIC formalism
is already discussed in~%
\cite{2024LohmayerLynchLeyendecker}.

The presented approach to
expressing multibody systems
is closely related to
the work of Sonneville and Brüls~\cite{%
2014SonnevilleBruls,%
2015Sonneville%
},
in the sense that
essentially the same variables and constraints
are used.
The EPHS approach makes explicit
the tearing and interconnection,
based on which systems can be efficiently expressed.
While it might seem irrelevant for multibody systems as such,
the structural properties of EPHS demand that
mechanical friction in the joint is
modeled in a thermodynamically consistent way,
making it straightforward to
include thermal dynamics.

To a somewhat lesser extent,
our approach is related to
previous work on
the port-Hamiltonian modeling of multibody systems~%
\cite{%
2001Stramigioli,%
2008MacchelliMelchiorri%
}.

We demonstrate that
our EPHS models agree
with equations obtained from
a suitable first principle from mechanics.
Specifically,
we use the Lagrange-d'Alembert-Pontryagin principle
for implicit Lagrangian systems
with constraints and external forces~%
\cite{2006YoshimuraMarsden2}.
Since our models use velocity variables
that are expressed in a body-fixed reference frame,
we use the left-trivialized version of the principle,
see~\cite{2008RabeeMarsden}.
Moreover,
to model body and joint as separate subsystems,
we use the tearing and interconnection approach
for implicit Dirac-Lagrange systems~%
\cite{2014JacobsYoshimura}.
Finally,
to match the thermodynamically consistent description
of mechanical friction,
we use a constraint of thermodynamic type~%
\cite{2017GayYoshimura1,2020GayYoshimura}.

\subsection{Outline}%

\Cref{sec:ephs}
provides an introduction to
the EPHS language
and it
introduces
the proposed multibody framework
in terms of
interconnection patterns for
the body model,
the joint model,
and a basic multibody system.
\Cref{sec:geometry}
provides an introduction to
the geometric description of multibody systems.
\Cref{sec:body}
completes the definition of the body model
by specifying the involved primitive subsystems,
while
\cref{sec:joint} does the same for the joint model.
For both models,
it is shown that
the resulting evolution equations
agree with
the Lagrange-d'Alembert-Pontryagin principle
for interconnected implicit Lagrangian systems
(with a constraint of thermodynamic type).
\Cref{sec:mbs}
shows the system of differential-algebraic equations
resulting from
the interconnected EPHS and variational models.
\Cref{sec:discussion}
concludes with
a discussion of
the results
and
possible directions for further research.

\section{EPHS modeling language by example}%
\label{sec:ephs}

With a detailed and formally precise introduction
being available in~%
\cite{2024LohmayerLynchLeyendecker},
here we aim for
a brief example-driven introduction.
We start with an analogy.
The essence of functional programming is function composition.
Complex functions are implemented in terms of simpler functions,
which again are composed of yet simpler functions.
The functions on the lowest level are
primitives provided by the execution environment.
Analogously,
the essence of EPHS
is system interconnection.
Complex systems are specified in terms of simpler systems,
which themselves are composite systems.
On the lowest level,
primitive systems
are defined by geometric objects
that represent primitive physical behaviors,
namely energy storage
and reversible/irreversible energy exchange.

\begin{figure}[htb]
	\centering
  \includegraphics[width=5.1cm]{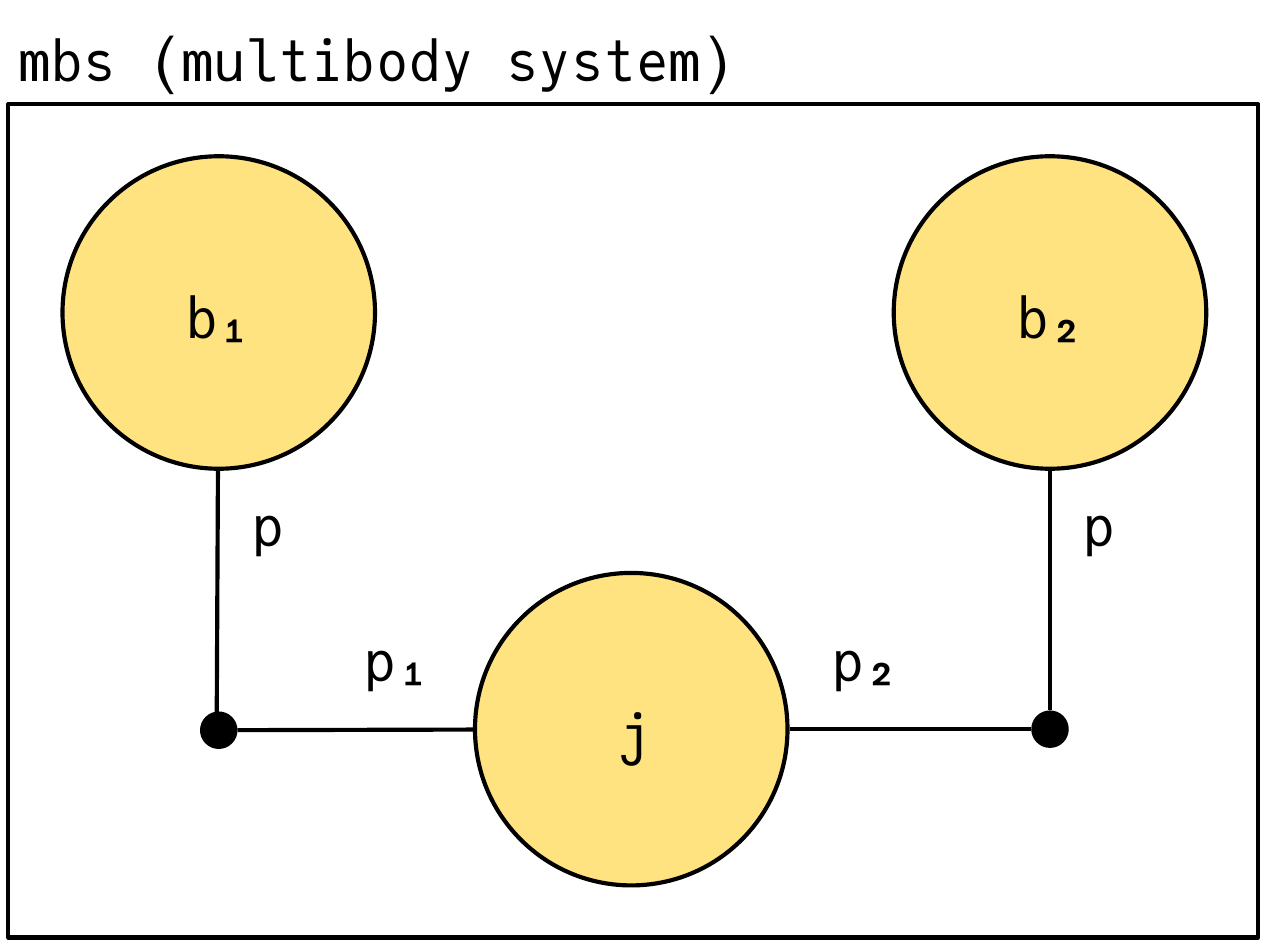}
  \caption{%
    Interconnection pattern for
    a basic multibody system
    consisting of
    two bodies $\mathtt{b_1}$ and $\mathtt{b_2}$
    connected by a joint $\mathtt{j}$.
  }%
	\label{fig:mbs}
\end{figure}

While a graphical syntax
for function composition
is directed
(similar to block diagrams),
the EPHS syntax is undirected.
Rather than feeding outputs to inputs,
it expresses how various systems
share energy domains.
For instance,
the pattern shown in~\cref{fig:mbs}
defines a composite system
in terms of three given subsystems.
The name $\mathtt{mbs}$
is analogous to a function name,
while
$\mathtt{b_1}$, $\mathtt{b_2}$ and $\mathtt{j}$
are analogous to argument names.
Specifically,
boxes $\mathtt{b_1}$ and $\mathtt{b_2}$
represent two rigid bodies
and $\mathtt{j}$ represents a joint that connects them.
At this level,
extending the multibody system to include more bodies
and joints becomes conceptually very simple.
The labor essentially reduces to
setting the parameters associated to
the subsystems in a meaningful way.
The lines emanating from the inner boxes are inner ports
which expose energy domains of the subsystems.
Here,
the ports $\mathtt{b_1.p}$ and $\mathtt{b_2.p}$
expose the kinetic energy domain of the two bodies.
These energy domains are shared with
the joint system $\mathtt{j}$,
which essentially applies constraint forces
such that the joint kinematics are satisfied.
The rectangular outer box represents
the interface of the composite system.
Here, there are no outer ports,
so the composite system is isolated.

The round inner boxes as well as the outer box
represent interfaces,
which are collections of ports.
Although not shown in the graphical representation,
next to its name,
each port is defined by a physical quantity,
which is the quantity that can be exchanged via the port.
In~\cref{fig:mbs} all ports
belong to a kinetic energy domain,
so momentum is the exchanged physical quantity.
All quantities have an associated space of values.
This extra data ensures that interconnections are well typed.
In summary,
each box represents the interface of a system.
An interface is given by
a collection of ports with associated quantities.
Each port exposes an energy domain
of the system to which it belongs.
Whenever two ports are connected at a junction,
the respective systems
may share information about the current value of the respective quantity
and they may exchange energy by exchanging the respective quantity.

\begin{figure}[htb]
  \centering
	\includegraphics[width=8.8cm]{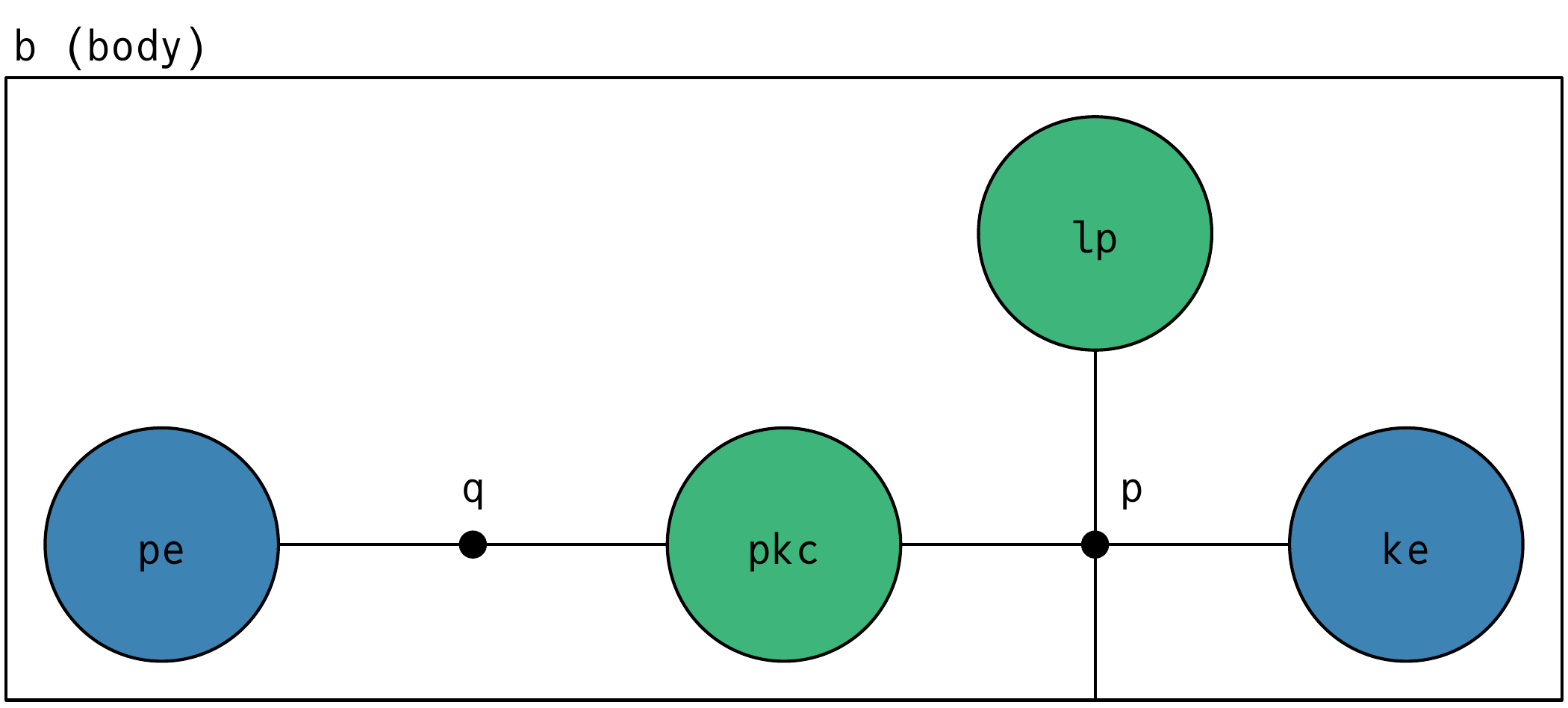}
  \caption{%
    Interconnection pattern for the rigid body model.
    Box $\mathtt{pe}$ represents
    storage of potential energy,
    while box $\mathtt{ke}$ represents
    storage of kinetic energy.
    Box $\mathtt{pkc}$ represents
    the reversible coupling
    of the potential and kinetic energy domains.
    Box $\mathtt{lp}$ represents the
    gyroscopic effects (Lie-Poisson structure).
  }%
	\label{fig:body}
\end{figure}

\begin{figure}[htb]
	\centering
	\includegraphics[width=5.2cm]{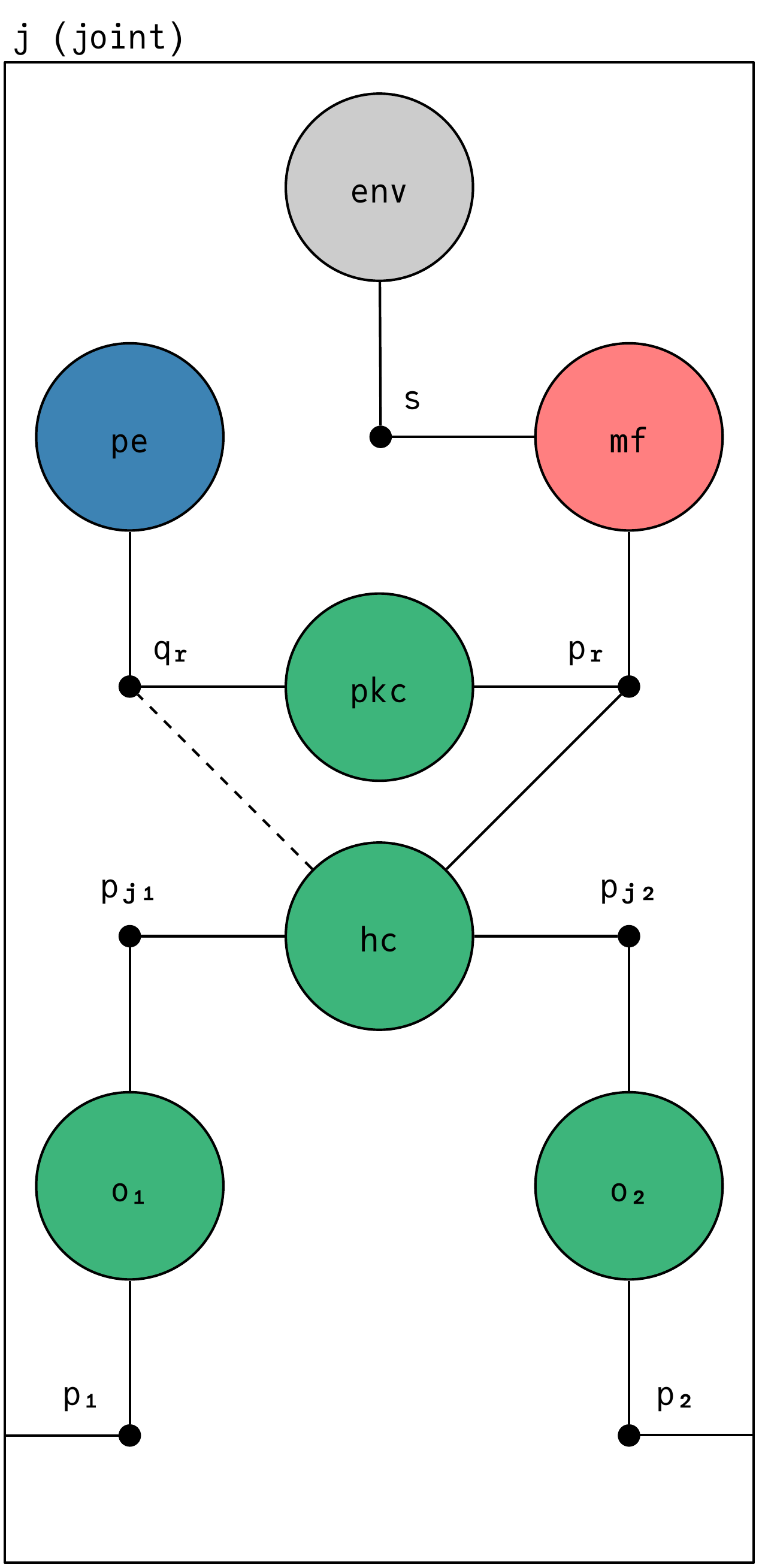}
  \caption{%
    Interconnection pattern for the joint model.
    Box $\mathtt{hc}$ represents the holonomic constraint
    describing the joint kinematics.
    Boxes $\mathtt{o_1}$ and $\mathtt{o_2}$
    take into account the
    offset between
    the reference frame of one of the connected bodies
    and
    the reference frame defining the respective joint force application point.
    Box $\mathtt{pe}$ represents a possible storage of
    potential energy, depending on the relative pose
    of the two connected bodies.
    Box $\mathtt{pkc}$ represents the coupling between
    the potential and kinetic energy domains
    of the joint.
    Box $\mathtt{mf}$ represents
    the irreversible process of mechanical friction
    and
    box $\mathtt{env}$ represents the environment
    which directly absorbs the generated heat.
  }%
	\label{fig:joint}
\end{figure}

\begin{figure}[htb]
	\centering
	\includegraphics[width=\textwidth]{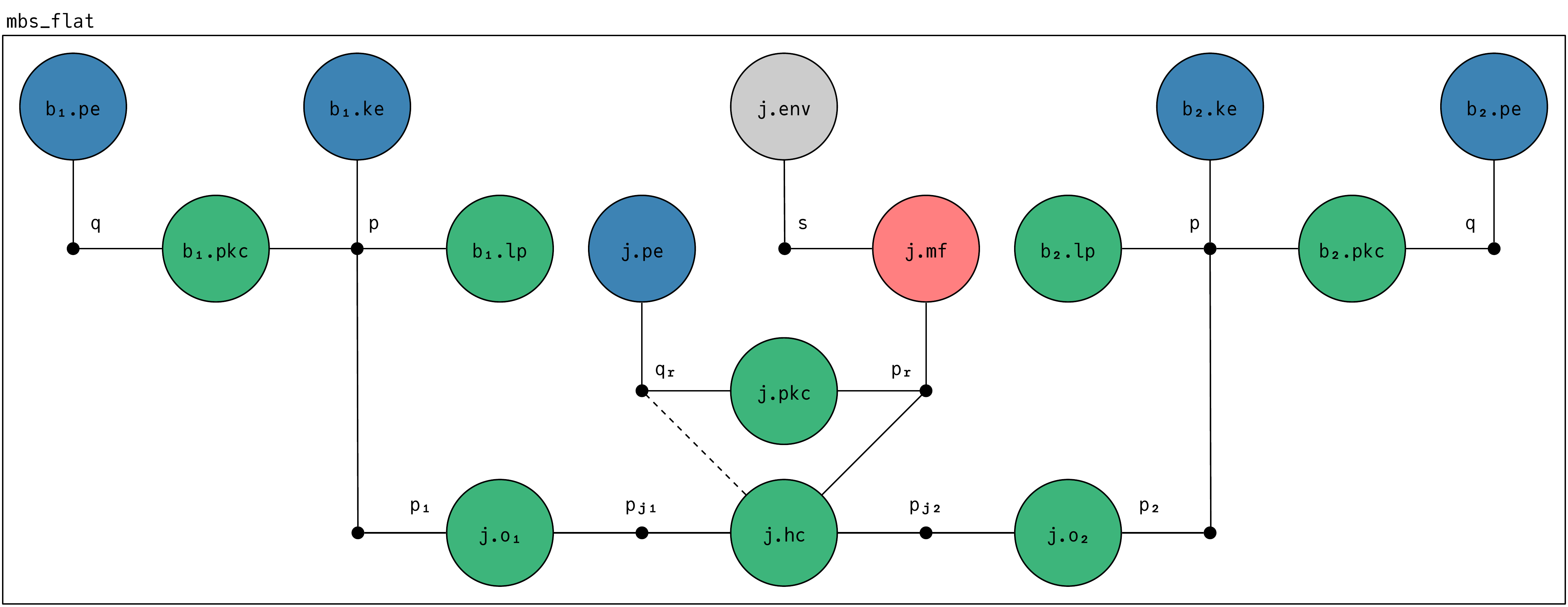}
  \caption{%
    Flattened interconnection pattern for
    the basic multibody system.
    The pattern is obtained by
    substituting the pattern in~\cref{fig:body}
    into the inner boxes $\mathtt{b_1}$ and $\mathtt{b_2}$
    of the pattern in~\cref{fig:mbs}
    and further by
    substituting the pattern in~\cref{fig:joint}
    into the inner box $\mathtt{j}$.
  }%
	\label{fig:mbs_flat}
\end{figure}

The yellow color of the inner boxes
in~\cref{fig:mbs}
is an annotation indicating that
the body and joint subsystems
are again composite systems
and therefore defined
via separate interconnection patterns.
The patterns for the body and joint models
are shown in~\cref{fig:body,fig:joint}, respectively.
It is important to note that
the hierarchical nesting of systems is defined since
the interface represented by the outer box
of the body/joint pattern
is equivalent (up to a renaming of ports) to
the interface represented by the respective inner boxes
in~\cref{fig:mbs}.
To illustrate this,
the flattened pattern is
shown explicitly in~\cref{fig:mbs_flat}.
Based on this uniquely defined notion of composition,
systems can be easily and safely decomposed
(or refactored)
into manageable and reusable parts.

We briefly remark that
there are two kinds of ports.
The dashed line in~\cref{fig:joint}
represents a state port,
rather than a power port.
A state port may exchange information
about the state of the respective quantity,
while a power port additionally allows for energy exchange.

We also remark that
we use an abbreviated notation to write down
the port names when visualizing interconnection patterns.
Whenever multiple ports connected to the same junction
have the same name,
we write the name only once at the junction.

\begin{figure}[htb]
  \centering
	\includegraphics[width=8.8cm]{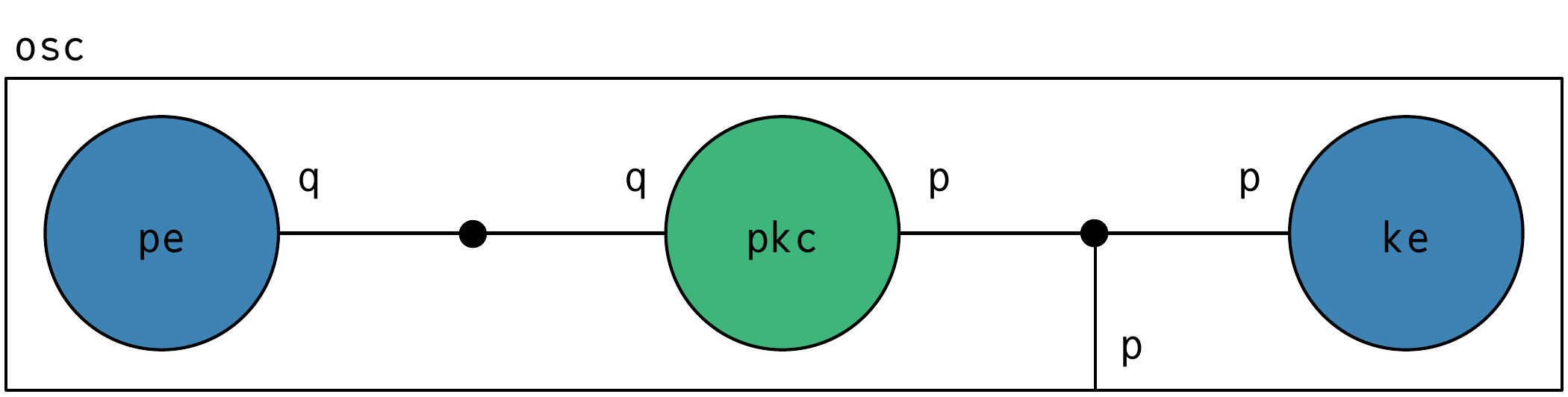}
  \caption{%
    Interconnection pattern for
    a mechanical oscillator model.
    Box $\mathtt{pe}$
    represents storage of potential energy.
    Box $\mathtt{ke}$
    represents storage of kinetic energy.
    Box $\mathtt{pkc}$
    represents the reversible coupling
    between the potential energy domain,
    represented by the junction on its left,
    and the kinetic energy domain
    represented by the junction on its right.
    The outer port $\mathtt{p}$
    exposes the kinetic energy domain.
  }%
	\label{fig:osc}
\end{figure}

To go into more detail,
we now turn our attention to the simplest example,
namely that of a 1D mechanical oscillator,
which is defined using
the interconnection pattern shown in~\cref{fig:osc}.
The pattern is similar to
that of the rigid body model,
except
there is no subsystem
for gyroscopic effects.
Moreover,
the interfaces of
the equally-named subsystems are not equivalent,
since the quantities of the corresponding ports
have different underlying spaces.
Here, they are just $1$-dimensional.
Next,
we define the three primitive subsystems
including their interfaces
and then
we come back to
the interconnection pattern
and show how the three systems are combined according to it.

When we define interfaces,
we need to choose quantities for the ports
form a set of possible quantities $\QuantitySet$.
Rather than defining this set upfront,
we simply state the relevant elements (quantities) as we go.
Besides that,
systems are defined with respect to
an exergy reference environment,
which enables the thermodynamically consistent combination
of reversible and irreversible dynamics.
For the purposes of this paper,
it is sufficient to define
the absolute environment temperature
${\color{violet} \theta}$.

We now define the
first primitive system,
namely the storage component
$(I_\text{pe}, \, E_\text{pe})$
filling the box $\mathtt{pe}$.
Its interface
$
I_\text{pe} =
(\{ \mathtt{q} \}, \, \tau_\text{pe})
$ is defined by
the set of ports
$\{ \mathtt{q} \}$
and a function
$
\tau_\text{pe} :
\{ \mathtt{q} \} \rightarrow
\mathbf{Q} \times \{ \mathsf{p}, \, \mathsf{s} \}
$
that assigns to each port
its quantity
as well as
a Boolean value indicating
whether the port is a power or a state port.
Here, we have
$
\tau_\text{pe}(\mathtt{q}) =
((\bR, \, \mathtt{displacement}), \, \mathsf{p})
$,
where
$(\bR, \, \mathtt{displacement}) \in \QuantitySet$
represents
the quantity displacement
with underlying state space $\bR$,
while the choice
$\mathsf{p} \in \{ \mathsf{p}, \, \mathsf{s} \}$
indicates that $\mathtt{q}$ is a power port.
The state space $\cX_{I}$
associated to an interface $I$
is the Cartesian product of
the state spaces of its ports.
Here, we simply have
$\cX_{I_\text{pe}} = \bR$.
Next to its interface,
a storage component is defined by
an energy function,
which assigns to each state
the corresponding stored energy.
Here, we assume a Hookean spring.
Hence,
the function
$
E_\text{pe} : \cX_{I_\text{pe}} \rightarrow \bR
$
is defined by
$
E_\text{pe}(q) = \frac{1}{2} \, k \, q^2
$,
where
$q \in \cX_{I_\text{pe}}$
denotes the displacement
and
$k \in \bR$ is a parameter of the model (stiffness).

All ports have
a state variable $\mathtt{x}$,
while power ports additionally have
a flow variable $\mathtt{f}$
and an effort variable $\mathtt{e}$.
For the power port $\mathtt{q}$,
we have
the port variables
$
(\mathtt{q.x}, \, \mathtt{q.f}, \, \mathtt{q.e})
\in \mathrm{T} \cX_\mathtt{q} \oplus \mathrm{T}^* \cX_\mathtt{q}
$,
where $\cX_\mathtt{q} = \bR$
is the state space of the port
and
$\mathrm{T} \cX_\mathtt{q} \oplus \mathrm{T}^* \cX_\mathtt{q}$
denotes the Whitney sum of
the tangent bundle and cotangent bundle over $\cX_\mathtt{q}$.
These geometric concepts are
briefly discussed in~\cref{sec:geometry}.
In this case,
we can simply identify
$
\mathrm{T} \cX_\mathtt{q} \oplus \mathrm{T}^* \cX_\mathtt{q}
\cong
\bR \times \bR \times \bR
$.
All port variables of an interface
together form a vector bundle.
Based on this,
the semantics of
interconnection patterns,
primitive systems,
and composite systems
can be understood geometrically
within a simple framework based on
relations between such bundles
and the composition of these relations~%
\cite{2024LohmayerLynchLeyendecker}.
Here, we discuss semantics
in terms of equations only.

The semantics of
the storage component
$(I_\text{pe}, \, E_\text{pe})$
filling the box $\mathtt{pe}$
is given by
\begin{equation}
  \begin{split}
    \mathtt{pe.q.x}
    \: &= \:
    q
    \\
    \mathtt{pe.q.f}
    \: &= \:
    \dot{q}
    \\
    \mathtt{pe.q.e}
    \: &= \:
    \dd E_\text{pe}(q)
    \: = \:
    k \, q
    \,.
  \end{split}
  \label{eq:osc_pe}
\end{equation}
The flow variable
is the rate of change
of the state variable (velocity)
and the effort variable
is the differential of the stored energy
at the current state (force).
The duality pairing
$
\langle \mathtt{q.e} \mid \mathtt{q.f} \rangle =
\langle \dd E_\text{pe}(q) \mid \dot{q} \rangle =
k \, q \, \dot{q}
$
is the power supplied to the system.
The concepts of differential and linear duality
are discussed in~\cref{sec:geometry}.
With the identification
$
\mathrm{T} \cX_\mathtt{q} \oplus \mathrm{T}^* \cX_\mathtt{q}
\cong
\bR \times \bR \times \bR
$,
the duality pairing simply becomes scalar multiplication.

We also remark that
in general
the effort variables are given by
the differential of
an exergy function
that is induced from
the energy function
based on the reference environment.
Since all storage components
in this paper
store forms of mechanical energy,
the respective
exergy and energy functions are equal.

The storage component
$(I_\text{ke}, \, E_\text{ke})$
filling the box $\mathtt{ke}$
is defined by
its interface
$
I_\text{ke} =
(\{ \mathtt{p} \}, \, \tau_\text{ke})
$
with
$
\tau_\text{ke}(\mathtt{p}) =
((\bR, \, \mathtt{momentum}), \, \mathsf{p})
$
and
its energy function
$
E_\text{ke} : \cX_{I_\text{ke}} \rightarrow \bR
$
given by
$
E_\text{ke}(p) = \frac{1}{2 \, m} \, p^2
$,
where
$p \in \cX_{I_\text{ke}} = \bR$
denotes the momentum
and
$m \in \bR$ is a parameter of the model (mass).
The semantics is hence given by
\begin{equation}
  \begin{split}
    \mathtt{ke.p.x}
    \: &= \:
    p
    \\
    \mathtt{ke.p.f}
    \: &= \:
    \dot{p}
    \\
    \mathtt{ke.p.e}
    \: &= \:
    \dd E_\text{ke}(p)
    \: = \:
    \frac{1}{m} \, p
    \,.
  \end{split}
  \label{eq:osc_ke}
\end{equation}

The reversible component
$(I_\text{pkc}, \, \cD_\text{pkc})$
filling the box $\mathtt{pkc}$
is defined by
its interface
$
I_\text{pkc} =
(\{ \mathtt{q}, \, \mathtt{p} \}, \, \tau_\text{pkc})
$
with
$
\tau_\text{pkc}(\mathtt{q}) =
((\bR, \, \mathtt{displacement}), \, \mathsf{p})
$
as well as
$
\tau_\text{pkc}(\mathtt{p}) =
((\bR, \, \mathtt{momentum}), \, \mathsf{p})
$
and its Dirac structure $\cD_\text{pkc}$,
which defines the semantics
given by
\begin{equation}
  \left[
    \begin{array}{c}
      \mathtt{pkc.q.f} \\
      \mathtt{pkc.p.f}
    \end{array}
  \right]
  \: = \:
  \left[
    \begin{array}{rrr}
      0 & -1 \\
      1 &  0
    \end{array}
  \right]
  \,
  \left[
    \begin{array}{c}
      \mathtt{pkc.q.e} \\
      \mathtt{pkc.p.e}
    \end{array}
  \right]
  \,.
  \label{eq:osc_pkc}
\end{equation}
We note that
the skew-symmetric matrix
used to represent a Dirac structure
implies conservation of power.
Here,
we have
$
\langle \mathtt{q.e} \mid \mathtt{q.f} \rangle
+
\langle \mathtt{p.e} \mid \mathtt{p.f} \rangle
= 0
$.

Coming back to the interconnection pattern
in~\cref{fig:osc},
the semantics is given
junction by junction.
We have
\begin{subequations}
  \begin{equation}
    \begin{alignedat}{2}
      &\mathtt{pe.q.x} \: {}={} \: &&\mathtt{pkc.q.x}
      \\
      &\mathtt{pe.q.f} \: {}+{} \: &&\mathtt{pkc.q.f} \: {}={} \: 0
      \\
      &\mathtt{pe.q.e} \: {}={} \: &&\mathtt{pkc.q.e}
    \end{alignedat}
  \end{equation}
  and
  \begin{equation}
    \begin{alignedat}{3}
      &\mathtt{pkc.p.x} \: {}={} \: &&\mathtt{ke.p.x} \: {}={} \: \mathtt{p.x}
      \\
      &\mathtt{pkc.p.f} \: {}+{} \: &&\mathtt{ke.p.f} \: {}={} \: \mathtt{p.f}
      \\
      &\mathtt{pkc.p.e} \: {}={} \: &&\mathtt{ke.p.e} \: {}={} \: \mathtt{p.e}
      \,.
    \end{alignedat}
  \end{equation}
  \label{eq:osc_pattern}%
\end{subequations}
In general,
at every junction
the state variables of all connected ports are equal.
Further,
the sum of the flow variables of all connected inner power ports
is equal to
the sum of the flow variables of all connected outer power ports.
%
Finally,
the effort variables of all connected power ports are equal.
By eliminating interface variables,
\cref{eq:osc_pe,eq:osc_ke,eq:osc_pkc,eq:osc_pattern}
can be reduced to
\begin{equation}
  \begin{split}
    \dot{q}
    \: &= \:
    \frac{1}{m} \, p
    \: = \:
    \mathtt{p.e}
    \\
    \dot{p}
    \: &= \:
    -k \, q \, + \, \mathtt{p.f}
    \,.
  \end{split}
  \label{eq:osc}
\end{equation}

\begin{figure}[htb]
  \centering
	\includegraphics[width=8.8cm]{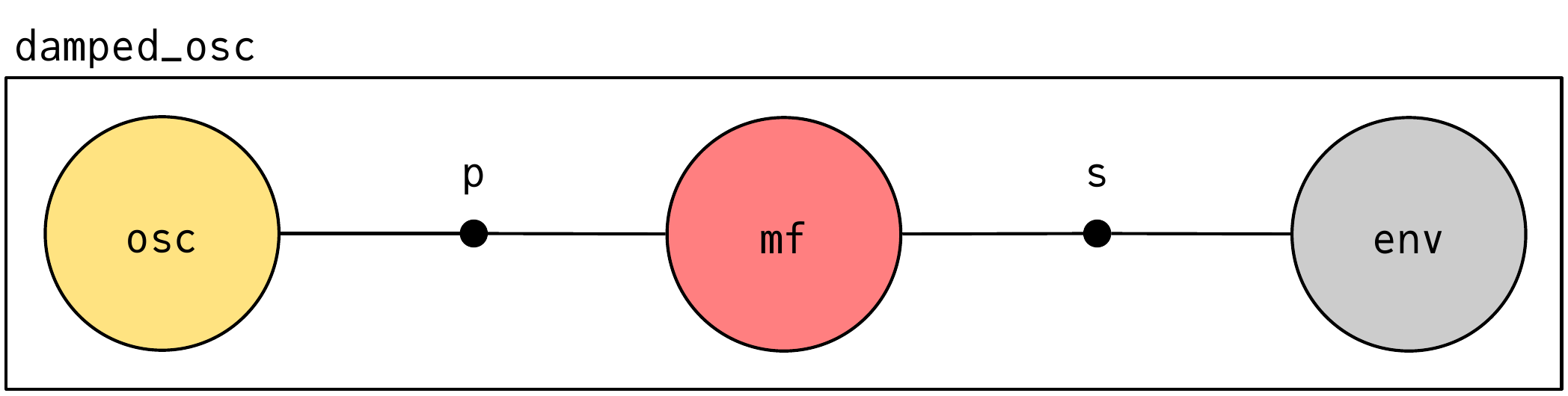}
  \caption{%
    Interconnection pattern for a damped oscillator model.
    Box $\mathtt{osc}$ represents
    the mechanical oscillator model
    defined based on the pattern in~\cref{fig:osc}.
    Box $\mathtt{mf}$ represents mechanical friction
    and
    box $\mathtt{env}$ represents the environment,
    which absorbs the generated heat.
  }%
	\label{fig:damped_osc}
\end{figure}

To add an irreversible process to the system,
we regard it as a subsystem
filling the inner box $\mathtt{osc}$
of the interconnection pattern
shown in~\cref{fig:damped_osc}.
Next we define the two additional components.

The irreversible component
$(I_\text{mf}, \, \cO_\text{mf})$
filling the box $\mathtt{mf}$
is defined by
its interface
$
I_\text{mf} =
(\{ \mathtt{p}, \, \mathtt{s} \}, \, \tau_\text{mf})
$
with
$
\tau_\text{mf}(\mathtt{p}) =
((\bR, \, \mathtt{momentum}), \, \mathsf{p})
$,
$
\tau_\text{mf}(\mathtt{s}) =
((\bR, \, \mathtt{entropy}), \, \mathsf{p})
$
and its Onsager structure $\cO_\text{mf}$,
which defines the semantics given by
\begin{equation}
  \begin{bmatrix}
    \mathtt{mf.p.f} \\
    \mathtt{mf.s.f}
  \end{bmatrix}
  \: = \:
  \frac{1}{\textcolor{violet}{\theta_0}} \,
  d \,
  \begin{bmatrix}
    \theta & -\upsilon \\
    -\upsilon & \frac{\upsilon^2}{\theta}
  \end{bmatrix}
  \,
  \begin{bmatrix}
    \mathtt{mf.p.e} \\
    \mathtt{mf.s.e}
  \end{bmatrix}
  \: = \:
  \begin{bmatrix}
    d \, \upsilon \\
   -\frac{1}{\theta} \, d \, \upsilon^2
  \end{bmatrix}
  \,.
  \label{eq:osc_mf}
\end{equation}
Here,
$\upsilon = \mathtt{p.e}$
is the velocity
and
$\theta = \textcolor{violet}{\theta_0} + \mathtt{s.e}$
is the absolute temperature
at which kinetic energy is dissipated
into the thermal energy domain.
These variables can be interpreted as
thermodynamic driving forces.
Further,
$d \in \bR$ is a parameter (friction coefficient).
The flow variable
$\mathtt{p.f}$
is the friction force
and
$\mathtt{s.f}$
is the entropy production rate,
with an additional minus sign,
since entropy leaves the system.
These variables can be interpreted as
the resulting thermodynamic fluxes.
The symmetric non-negative definite matrix
used to define the Onsager structure
implies non-negative exergy destruction,
which is equivalent to non-negative entropy production.
Specifically,
the exergy destruction rate is
$
\langle \mathtt{p.e} \mid \mathtt{p.f} \rangle
+
\langle \mathtt{s.e} \mid \mathtt{s.f} \rangle
=
\textcolor{violet}{\theta_0} \,
\frac{1}{\theta} \, d \, \upsilon^2
$,
where
$\frac{1}{\theta} \, d \, \upsilon^2$
is the entropy production rate.
Energy is conserved since
(see~\cite{2024LohmayerLynchLeyendecker} for details)
\begin{equation*}
  \begin{bmatrix}
    \theta & -\upsilon \\
    -\upsilon & \frac{\upsilon^2}{\theta}
  \end{bmatrix}
  \,
  \begin{bmatrix}
    \upsilon \\
    \theta
  \end{bmatrix}
  \: = \:
  \begin{bmatrix}
   0 \\
   0
  \end{bmatrix}
  \,.
\end{equation*}

We note that
the Dirac/Onsager structures of
reversible/irreversible components
in general
have to satisfy further properties
to ensure thermodynamic consistency~%
\cite{2024LohmayerLynchLeyendecker}.

The environment component
essentially is a storage component that
represents the thermal energy domain of
the isothermal reference environment.
By definition,
its exergy function
always takes the value zero.
Hence,
the effort variable is also zero.
Its semantics is then given by
\begin{equation}
  \begin{split}
    \mathtt{env.s.x}
    \: &= \:
    s
    \\
    \mathtt{env.s.f}
    \: &= \:
    \dot{s}
    \\
    \mathtt{env.s.e}
    \: &= \:
    0
    \,.
  \end{split}
  \label{eq:osc_env}
\end{equation}

By eliminating interface variables,
\cref{eq:osc,eq:osc_mf,eq:osc_env}
combined with those for
the pattern in~\cref{fig:damped_osc}
can be reduced to
\begin{equation}
  \begin{split}
    \dot{q}
    \: &= \:
    \upsilon
    \\
    \dot{p}
    \: &= \:
    -k \, q - d \, \upsilon
    \\
    \dot{s}
    \: &= \:
    \frac{1}{\theta_0} \, d \, \upsilon^2
    \,,
  \end{split}
  \label{eq:damped_osc}
\end{equation}
where
$\upsilon = \frac{1}{m} \, p$.

\begin{figure}[htb]
  \centering
	\includegraphics[width=10.2cm]{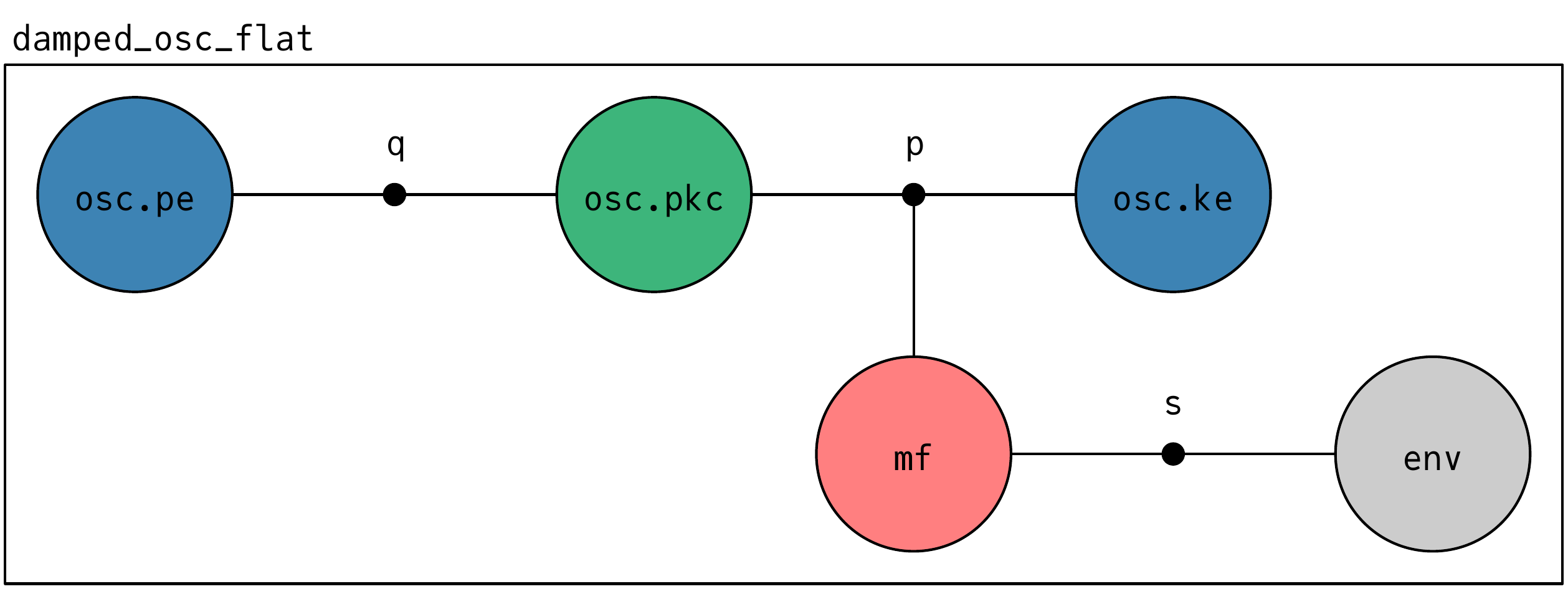}
  \caption{Interconnection pattern of the rigid body model.}%
	\label{fig:damped_osc_flat}
\end{figure}

We may say that the semantics of EPHS is functorial
since~\cref{eq:damped_osc}
can be equally obtained by
first substituting the pattern in~\cref{fig:osc}
into the pattern in~\cref{fig:damped_osc},
which gives the pattern in~\cref{fig:damped_osc_flat},
and then filling its inner boxes
with the same components.
Hence,
the composition of interconnection patterns (syntax)
is indeed compatible with interconnecting systems (semantics).

\section{Geometric foundation}%
\label{sec:geometry}

Here,
we aim for a brief
introduction
to the geometric concepts used
in this paper.
More details
can be found in
textbooks such as~%
\cite{2012Lee}
or~\cite{1999MarsdenRatiu,1988AbrahamMarsdenRatiu}.

\subsection{Physical space and coordinate frames}%

The \textbf{Euclidean vector space}
$(\mathbb{E}^3, \, \langle \cdot \ , \, \cdot \rangle)$
is a real $3$-dimensional vector space $\mathbb{E}^3$
with an inner product
$
\langle \cdot \ , \, \cdot \rangle :
\mathbb{E}^3 \times \mathbb{E}^3 \rightarrow \mathbb{R}
$.

A choice of orthonormal basis for $\mathbb{E}^3$
gives an isomorphism
$\mathbb{E}^3 \cong \mathbb{R}^3$.
The orthonormality condition ensures that
the isomorphism preserves the inner product,
with the inner product of $\mathbb{R}^3$
being the dot product.
Specifically,
let's
consider two right-handed orthonormal bases
$(e_1^I, \, e_2^I, \, e_3^I)$ and
$(e_1^B, \, e_2^B, \, e_3^B)$.
A vector $v \in \mathbb{E}^3$
can then be represented as
\begin{equation*}
  v
  \: = \:
  {}_Iv^1 \, e_1^I + {}_Iv^2 \, e_2^I + {}_Iv^3 \, e_3^I
  \: = \:
  {}_Bv^1 \, e_1^B + {}_Bv^2 \, e_2^B + {}_Bv^3 \, e_3^B
  \,,
\end{equation*}
where
${}_Iv = ({}_Iv^1, \, {}_Iv^2, \, {}_Iv^3) \in \mathbb{R}^3$
is the \textbf{coordinate representation} of $v$
with respect to basis $I$
and
${}_Bv = ({}_Bv^1, \, {}_Bv^2, \, {}_Bv^3) \in \mathbb{R}^3$
is its representation
with respect to basis $B$.
For computations in coordinates,
we regard these as column vectors,
i.e.~we identify
$\mathbb{R}^3 \cong \mathbb{R}^{3 \times 1}$,
and
we simply use juxtaposition
to denote matrix multiplication.
For instance,
we have
$
\langle v, \, v \rangle =
{}_Iv^\mathrm{T} {}_Iv =
{}_Bv^\mathrm{T} {}_Bv
$,
since $I$ and $B$ are orthonormal.

The two representations are related by
\begin{equation*}
  {}_Iv
  \: = \:
  R_{IB} \, {}_Bv
  \,,
\end{equation*}
where
the \textbf{coordinate transformation matrix}
$R_{IB} \in \mathbb{R}^{3 \times 3}$
contains the basis vectors of $B$
expressed in the basis $I$, i.e.
$R_{IB} = \bigl[ {}_Ie_1^B\ \mid {}_Ie_2^B\ \mid {}_Ie_3^B \bigr]$.
With $E \in \mathbb{R}^{3 \times 3}$
denoting the identity matrix,
we have
$R_{IB}^\mathrm{T} \, R_{IB} = R_{BI} \, R_{IB} = E$,
since the transformation preserves the inner product,
and moreover we have
$\det(R_{IB}) = 1$,
as it also preserves the right-handed orientation.
$R_{IB}$ is commonly called a rotation matrix.

The \textbf{Euclidian (affine) space}
$(\cE^3, \, +)$
with associated vector space $\bE^3$
shall represent physical space.
By definition,
for any two points $P, \, Q \in \cE^3$,
there is a unique vector
denoted by
$r_{PQ} \in \mathbb{E}^3$
that represents the \textbf{translation} from $P$ to $Q$.
So,
the vector space $\mathbb{E}^3$
acts (transitively) on
the point space $\cE^3$
by translation
and
we have
$P + r_{PQ} = Q$,
where
$+ : \cE^3 \times \mathbb{E}^3 \rightarrow \cE^3$
denotes the (affine) action.

A coordinate \textbf{frame}
$(A, \, (e_1^A, \, e_2^A, \, e_3^A))$
is defined by
an origin $A \in \cE^3$
and
a basis for $\mathbb{E}^3$.
Hence,
a frame gives
an isomorphism $\cE^3 \cong \mathbb{R}^3$.
We reuse the same symbol, here $A$,
to refer to a frame
or just to its origin or its basis.
In particular,
we use $I$ to denote
the \textbf{inertial reference frame}
$(I, \, (e_1^I, \, e_2^I, \, e_3^I))$
used throughout.

\subsection{Configuration of a rigid body}%

At rest,
a \textbf{rigid body} $(\cB, \, \rho)$
is characterized by
its set of material points (reference configuration)
$\cB \subset \cE^3$
and
its mass density function
$\rho : \cB \rightarrow \mathbb{R}_+$.
Let
$(B, \, (e_1^B, \, e_2^B, \, e_3^B))$
be a frame with origin $B \in \cB$.
Considering the possibility of motion,
we say that $B$ is a \textbf{body-fixed frame}
when it moves with the body.
Assuming this,
the time-dependent position
of an arbitrary particle $P \in \cB$
with respect to the reference frame $I$
can be written as
\begin{equation}
  {}_Ir_{IP}(t)
  \: = \:
  {}_Ir_{IB}(t) + R_{IB}(t) \, {}_Br_{BP}
  \,,
  \label{eq:particle_pos}
\end{equation}
where
${}_Br_{BP} \in \mathbb{R}^3$
is constant due to the body's rigidity.
Hence,
the time-dependent \textbf{configuration}
of the rigid body
is given by
\begin{equation}
  q_{IB}(t)
  \: = \:
  \bigl( R_{IB}(t), \, {}_Ir_{IB}(t) \bigr)
  \: \in \:
  \mathrm{SO}(3) \times \mathbb{R}^3
  \,.
  \label{eq:absolute_pose}
\end{equation}
We also say that
$q_{IB}(t)$ is the \textbf{absolute pose} of the body,
as it is given relative to
the distinguished reference frame $I$.
The set of matrices
\begin{equation*}
  \mathrm{SO}(3)
  \: = \:
  \Bigl\{
    R \in \mathbb{R}^{3 \times 3}
    \, \mid \,
    R^\mathrm{T} R = E
    \: ; \:
    \det(R) = 1
  \Bigr\}
\end{equation*}
forms a matrix Lie group,
as
it forms a smooth $3$-dimensional space (manifold)
as well as a group,
with
the composition operation
(matrix multiplication)
and
the inverse operation
(matrix transposition)
being smooth maps.
$\mathrm{SO}(3)$ is called
special orthogonal group
or simply \textbf{rotation group}.

\subsection{Lie group structure of the configuration space}%

The Lie group structure of
the space of all possible configurations,
gives rise to
mathematical operations
that are central to
the description of
rigid-body dynamics.

Next to
$\mathrm{SO}(3)$,
also $\mathbb{R}^3$ is a Lie group,
since vector spaces are Lie groups,
with the composition operation being vector addition
and the inverse operation being scalar multiplication by $-1$.
The configuration space
$\mathrm{SO}(3) \times \mathbb{R}^3$
then also forms a Lie group,
called the \textbf{direct product}.

Because $\mathrm{SO}(3)$ acts on $\mathbb{R}^3$
by matrix-vector multiplication,
the configuration space admits yet another group structure,
called the \textbf{semidirect product}.
This group is denoted by
$\mathrm{SO}(3) \ltimes \mathbb{R}^3 = \mathrm{SE}(3)$
and it is also called the special Euclidian group.

Since both Lie groups can be used
as a geometric basis for
describing rigid body dynamics~%
\cite{2011BrulsArnoldCardona,2014MuellerTerze},
we use the symbol $G$ to denote either of the two.
The \textbf{composition operation}
$G \times G \rightarrow G$,
which we denote simply by juxtaposition,
is defined by
\begin{equation*}
  \begin{aligned}
    G = \mathrm{SO}(3) \times \mathbb{R}^3 \colon \quad&
    (R_1, \, r_1) \, (R_2, \, r_2)
    \: = \:
    (R_1 \, R_2, \, r_1 + r_2)
    \\
    G = \mathrm{SO}(3) \ltimes \mathbb{R}^3 \colon \quad&
    (R_1, \, r_1) \, (R_2, \, r_2)
    \: = \:
    (R_1 \, R_2, \, r_1 + {\color{violet} R_1} \, r_2)
    \,.
  \end{aligned}
\end{equation*}
In both cases,
the \textbf{identity element} is
$e = (E, \, 0) \in G$.
The \textbf{inverse operation}
${(\cdot)}^{-1} : G \rightarrow G$
is then defined by
\begin{equation*}
  \begin{aligned}
    G = \mathrm{SO}(3) \times \mathbb{R}^3 \colon \quad&
    (R, \, r)^{-1}
    \: = \:
    (R^\mathrm{T}, -r)
    \\
    G = \mathrm{SO}(3) \ltimes \mathbb{R}^3 \colon \quad&
    (R, \, r)^{-1}
    \: = \:
    (R^\mathrm{T}, \, -{\color{violet} R^\mathrm{T}} \, r)
    \,.
  \end{aligned}
\end{equation*}

\subsection{Geometric interpretation of composition}%
\label{ssec:composition_interpretation}

The difference between the two group structures
becomes apparent
when
the absolute pose
$q_{IB} = (R_{IB}, \, {}_Ir_{IB})$,
as defined by~\cref{eq:absolute_pose},
is composed with
a relative pose $q_{BC}$,
where $C$ is the origin of another frame
$(C, \, (e_1^C, \, e_2^C, \, e_3^C))$.
Let
$q_{IC} = (R_{IC}, \, {}_Ir_{IC})$
denote the corresponding absolute pose.
The relative pose $q_{BC}$
is then defined by
$q_{IC} = q_{IB} \, q_{BC}$.
As we show next,
the different composition operations
imply different definitions for $q_{BC}$.

The case
$G = \mathrm{SO}(3) \times \mathbb{R}^3$
corresponds to the relative translation $r_{BC}$
being expressed in the inertial basis $I$, i.e.~%
$q_{BC} = (R_{BC}, \, {}_Ir_{BC})$,
since based on
\begin{equation*}
  q_{IC}
  \: = \:
  q_{IB} \, q_{BC}
  \: = \:
  (R_{IB}, \, {}_Ir_{IB}) \, (R_{BC}, \, {}_Ir_{BC})
  \: = \:
  (R_{IB} \, R_{BC}, \, {}_Ir_{IB} + {}_Ir_{BC})
  \,
\end{equation*}
we can write
the position of an arbitrary particle $P \in \cB$ as
\begin{equation*}
  \begin{aligned}
    {}_Ir_{IP}
    \: &= \:
    {}_Ir_{IC} + R_{IC} \, {}_Cr_{CP}
    \\
    \: &= \:
    {\color{violet} {}_Ir_{IB} + {}_Ir_{BC}} +
    {\color{violet} R_{IB} \, R_{BC}} \, {}_Cr_{CP}
    \,.
  \end{aligned}
\end{equation*}

The case
$G = \mathrm{SO}(3) \ltimes \mathbb{R}^3$,
corresponds to the relative translation $r_{BC}$
being expressed in the basis $B$, i.e.~%
$q_{BC} = (R_{BC}, \, {}_Br_{BC})$,
since based on
\begin{equation*}
  q_{IC}
  \: = \:
  q_{IB} \, q_{BC}
  \: = \:
  (R_{IB}, \, {}_Ir_{IB}) \, (R_{BC}, \, {}_Br_{BC})
  \: = \:
  (R_{IB} \, R_{BC}, \, {}_Ir_{IB} + R_{IB} \, {}_Br_{BC})
  \,
\end{equation*}
we can write
the position of an arbitrary particle $P \in \cB$ as
\begin{equation*}
  \begin{aligned}
    {}_Ir_{IP}
    \: &= \:
    {}_Ir_{IC} + R_{IC} \, {}_Cr_{CP}
    \\
    \: &= \:
    {\color{violet} {}_Ir_{IB} + R_{IB} \, {}_Br_{BC}} +
    {\color{violet} R_{IB} \, R_{BC}} \, {}_Cr_{CP}
    \,.
  \end{aligned}
\end{equation*}


\subsection{Material velocity}%

According to~\cref{eq:particle_pos},
the position
of any particle $P \in \cB$
can be written
in terms of the configuration
$
q_{IB}(t) =
\bigl( R_{IB}(t), \, {}_Ir_{IB}(t) \bigr)
\in G
$
of the body $\cB$,
which is seen to be
a smooth curve
$q_{IB} : \mathbb{I} \to G$
with time interval
$\mathbb{I} \subset \mathbb{R}$.
Since ${}_Br_{BP}$ is constant,
the velocity of the particle is given by
\begin{equation*}
  {}_I\dot{r}_{IP}(t)
  \: = \:
  {}_I\dot{r}_{IB}(t) + \dot{R}_{IB}(t) \, {}_Br_{BP}
  \,,
\end{equation*}
which depends solely on
the \textbf{material velocity}
$
\dot{q}_{IB}(t) =
\bigl( \dot{R}_{IB}(t), \, {}_I\dot{r}_{IB}(t) \bigr)
\in \mathrm{T}_{q_{IB}} G
$.
The vector space
$\mathrm{T}_{q_{IB}} G$
of all possible velocities
at configuration $q_{IB}$
is discussed next.

\subsection{The derivative (interlude)}%

A \textbf{smooth manifold} $M$
has a notion of neighborhoods around points.
By definition,
for every neighborhood (open set) $U \subset M$,
there is a smooth isomorphism
$x \colon U \to x(U)$
that takes any point in $U$
to its coordinate representation
in $x(U) \subseteq \mathbb{R}^n$,
where $n = \dim(M)$.
The smooth map $x$ is called
a \textbf{coordinate chart} on $U$.
While a single chart suffices for a flat space
such as $\cE^3$,
in general it takes multiple overlapping charts
to cover a manifold.

As the derivative is a local operator,
it can be computed numerically
on the chart level.
Staying on the coordinate-free level,
let
$f: M \to N$
be any smooth function
between two smooth manifolds.
The \textbf{derivative} of $f$
evaluated at point $p \in M$
is the linear function
\begin{equation*}
  \begin{split}
    \mathrm{T}_p f \colon
    \mathrm{T}_p M &\to \mathrm{T}_{f(p)} N
    \\
    v &\mapsto
    \dv{t} \, \Big\vert_{t=0} \, f(c(t))
    \,,
  \end{split}
\end{equation*}
where
$c : \mathbb{I} \to M$
with $\mathbb{R} \supseteq \mathbb{I} \ni 0$
is an arbitrary smooth curve on $M$
such that
$c(0) = p$ and $\dot{c}(0) = v$.
So,
the derivative 
of a smooth curve $c$
at some point (in time, here $0$)
is a vector $v \in \mathrm{T}_{c(0)} M$,
which is seen to be
tangent to the curve at that point.
The \textbf{tangent space}
over point $p$,
denoted as $\mathrm{T}_p M$,
is simply the vector space of
all possible tangent vectors at $p$,
considering all possible curves passing through $p$.
Finally,
the derivative $\mathrm{T} f$
of any smooth function $f$
is the linear function
that propagates tangent vectors along $f$.
Going one level up,
the disjoint union of all tangent spaces
$\mathrm{T} M = \sqcup_{p \in M} \mathrm{T}_p M$
forms again a smooth manifold,
called the \textbf{tangent bundle} over $M$.
We have
$\dim(\mathrm{T} M) = 2 \cdot \dim(M)$,
since for every point $p \in M$,
there are
$\dim(\mathrm{T}_p M) = \dim(M)$
directions for change.
So,
$\mathrm{T}$ sends
a manifold $M$ to its tangent bundle $\mathrm{T} M$
and it sends
a smooth map $f : M \to N$ between manifolds to
its derivative $\mathrm{T} f : \mathrm{T} M \to \mathrm{T} N$.
For any composite function $f = f_2 \circ f_1$,
it satisfies the chain rule
(functor property)
$\mathrm{T} f = \mathrm{T} f_2 \circ \mathrm{T} f_1$.

We now connect this to
the \textbf{differential} $\dd f$
of a function $f \colon M \to \bR$,
as used in~\cref{eq:osc_pe,eq:osc_ke}.
First, we notice that
for a function $f \colon M \to N$
between arbitrary manifolds,
the derivative is a map
$\mathrm{T} f \colon \mathrm{T} M \to \mathrm{T} N$
that sends any pair
$(p, \, v)$
with $p \in M$ and $v \in \mathrm{T}_p M$
to the pair
$(q, \, w)$
with $q = f(p)$ and $w \in \mathrm{T}_{q} N$
being
the infinitesimal change of $f(p)$
when the local change of $p$ is given by $v$.
To make composition work (chain rule),
it is important to also propagate the local points
such as
$q$ and $p$
and not only the local changes
$v$ and $w$.
In contrast to the derivative,
the differential is a concept that
only applies to
functions $f : M \to \bR$
\emph{on} manifolds.
At point $p$,
the differential $\dd f(p)$
does not include the value $f(p)$.
It simply is a covector
$\dd f(p) \in \mathrm{T}^*_p M$,
i.e.~a linear function
that sends
a vector $v \in \mathrm{T}_p M$
to the corresponding infinitesimal change
$w \in \mathrm{T}_{f(p)} \bR \cong \bR$.
We write
$w = \langle \dd f(p) \mid v \rangle$
for the duality pairing.
The concept of covectors and linear duality
is discussed next.

\subsection{Dual spaces and dual maps (another interlude)}%

Given a vector space $V$,
its \textbf{dual space} $V^*$
is the vector space of linear functions
from $V$ to $\bR$.
Vector addition on $V^*$ is defined by
$
(\alpha_1 + \alpha_2)(v) =
\alpha_1(v) + \alpha_2(v)
$
for
any two dual vectors (covectors)
$\alpha_1, \, \alpha_2 \in V^*$
and
any vector $v \in V$.
Scalar multiplication on $V^*$
is also inherited from $\bR$, i.e.~%
$
(c \cdot \alpha)(v) =
c \cdot \alpha(v)
$
for
any covector $\alpha \in V^*$,
any scalar $c \in \bR$,
and
any vector $v \in V$.
The \textbf{duality pairing}
$
\langle \cdot \mid \cdot \rangle \colon
V^* \times V \to \bR
$
is simply defined by
$\langle \alpha \mid v \rangle = \alpha(v)$
for
any covector $\alpha \in V^*$
and
any vector $v \in V$.
A basis
$(e_1, \, \ldots, \, e_n)$ for $V$
determines
the corresponding dual basis
$(e^1, \, \ldots, \, e^n)$ for $V^*$
by requiring
$\langle e^i \mid e_j \rangle = \delta^i_j$
for all $i, \, j = 1, \, \ldots, \, n$,
where
$n = \dim(V) = \dim(V^*)$
and
$\delta^i_j = 1$ if $i=j$ and $\delta^i_j = 0$ otherwise.

Given a linear map $f \colon V \to W$
between two vector spaces $V$ and $W$,
the \textbf{dual map} $f^* \colon W^* \to V^*$
is defined by
$
\langle f^*(\alpha) \mid v \rangle =
\langle \alpha \mid f(v) \rangle
$
for any $v \in V$ and $\alpha \in W^*$.
Assuming a choice of basis for both $V$ and $W$,
linear maps $V \to W$
can be represented as matrices.
The matrix for $f^*$ then simply is
the transpose of the matrix for $f$.

Given a manifold $M$,
we can define
its \textbf{cotangent bundle}
$\mathrm{T}^* M = \sqcup_{p \in M} \mathrm{T}^*_p M$,
where
the cotangent space
$\mathrm{T}^*_p M$
is the dual space of
the tangent space
$\mathrm{T}_p M$.
%
%
%
The derivative
$\mathrm{T} f \colon \mathrm{T} M \to \mathrm{T} N$
of a function
$f \colon M \to N$
is an instance of a vector bundle map,
since for every point $p \in M$,
it provides a linear map from
$\mathrm{T}_p M$
to
$\mathrm{T}_{q} N$,
were $q = f(p)$.
Given a vector bundle map
$g \colon \mathrm{T} M \to \mathrm{T} N$
with underlying base map
$f \colon M \to N$,
its dual map
$g^* \colon \mathrm{T}^* N \to \mathrm{T}^* M$
is defined by
$
\langle g^*(\alpha) \mid v \rangle =
\langle \alpha \mid g(v) \rangle
$
for
any point $p \in M$,
any tangent vector $v \in \mathrm{T}_p M$ and
any cotangent vector $\alpha \in \mathrm{T}_{f(p)}^* N$.

The \textbf{Whitney sum}
$\mathrm{T} M \oplus \mathrm{T}^* M$
of
the tangent bundle $\mathrm{T} M$ and
the cotangent bundle $\mathrm{T}^* M$
has elements
$
(p, \, v_p, \, \alpha_p)
\in \mathrm{T} M \oplus \mathrm{T}^* M
$,
where
$p \in M$,
$v_p \in \mathrm{T}_p M$ and
$\alpha_p \in \mathrm{T}^*_p M$.

\subsection{Trivialized velocity variables}%
\label{ssec:left_trivialization}

Since
the motion of a rigid body
is described
as a curve on $G$,
a velocity variable naturally is
a vector in
the $12$-dimensional tangent bundle $\mathrm{T} G$.
Based on the group structure,
we can represent velocities
in a $6$-dimensional vector space.

First,
we note that
for any group $G$
and any element $\bar{q} \in G$,
there are
two canonical isomorphisms $G \rightarrow G$,
called \textbf{left and right translation}.
Left translation by $q$
is simply defined by
$L_{q}(\bar{q}) = q \, \bar{q}$,
while right translation by $q$
is defined by
$R_{q}(\bar{q}) = \bar{q} \, q$.
Let $f$ denote either of the two maps.

Since
$G$ is a Lie group,
$f \colon G \rightarrow G$
is a smooth isomorphism,
and
so its derivative
evaluated at any point $\bar{q} \in G$ is
a linear isomorphism
denoted by
$
\mathrm{T}_{\bar{q}} f \colon
\mathrm{T}_{\bar{q}} G \to \mathrm{T}_{f(\bar{q})} G
$.
Based on this,
we can push
any tangent vector
$v \in \mathrm{T}_{\bar{q}} G$
forward to
a single tangent space,
namely the tangent space
$\mathfrak{g} = \mathrm{T}_e G$
over the identity element $e$.
For this, we simply choose
$q = \bar{q}^{-1}$,
since
$L_{q^{-1}}(q) = R_{q^{-1}}(q) = e$.

Considering all $q \in G$,
the family of maps
$
\mathrm{T}_q L_{q^{-1}} \colon
\mathrm{T}_q G \rightarrow \mathfrak{g}
$
gives a vector bundle isomorphism
$\mathrm{T} G \cong G \times \mathfrak{g}$,
called \textbf{left-trivialization} of $\mathrm{T} G$.
So,
instead of
the material velocity
$\dot{q} \in \mathrm{T}_q G$,
we can use
the left-trivialized velocity variable
\begin{equation}
  \tilde{u}
  \: = \:
  \mathrm{T}_q L_{q^{-1}}(\dot{q})
  \: \in \: \mathfrak{g}
  \,.
\end{equation}
The right-trivialization of $\mathrm{T} G$
and
the right-trivialized velocity
is defined analogously.

To work out the lower-level details,
we first characterize
the vector space $\mathfrak{g}$.
While $R \in \mathrm{SO}(3)$ is
a $3 \times 3$ matrix,
we have
$\dim(\mathrm{SO}(3)) = 3$
due to the orthonormality constraint.
Let
$R \colon \bI \rightarrow \mathrm{SO}(3)$
be a curve
with $R(0) = E$.
Taking the derivative of
the orthonormality constraint
at $t = 0$
gives
\begin{equation*}
  \dv{t} \, \Big\vert_{t=0} \, R^\mathrm{T}(t) \, R(t)
  \: = \:
  \dot{R}^\mathrm{T}(0) \, R(0) +
  R^\mathrm{T}(0) \, \dot{R}(0)
  \: = \:
  \dot{R}^\mathrm{T}(0) + \dot{R}(0)
  \: = \:
  \dv{t} \, \Big\vert_{t=0} \, E
  \: = \:
  0
  \,.
\end{equation*}
Since
$\dot{R}(0) \in \mathrm{T}_E \mathrm{SO}(3)$,
we conclude that
\begin{equation*}
  \mathfrak{so}(3)
  \: \coloneqq \:
  \mathrm{T}_E \mathrm{SO}(3)
  \: = \:
  \Bigl\{
    \tilde{\omega} \in \mathbb{R}^{3 \times 3}
    \, \mid \,
    \tilde{\omega}^\mathrm{T} + \tilde{\omega} \, = \, 0
  \Bigr\}
  \,.
\end{equation*}
The vector space $\mathfrak{so}(3)$
of skew-symmetric $3 \times 3$ matrices
can be identified with $\bR^3$
by collecting
the three non-zero entries of
$\tilde{\omega} \in \mathfrak{so}(3)$
in a vector
$\omega \in \bR^3$
such that
for all $v \in \bR^3$
we have
$\tilde{\omega} \, v = \omega \times v$.
We denote
the isomorphism
$\mathfrak{so}(3) \cong \bR^3$
simply by the presence or absence of the tilde symbol.
Based on this
and the identification
$\mathrm{T}_0 \bR^3 \cong \mathbb{R}^3$,
we have
$
\mathfrak{g} \coloneqq
\mathrm{T}_e G =
\mathfrak{so}(3) \times \bR^3
$
and
we accordingly identify
$
\mathfrak{g} \ni \tilde{u}
=
(\tilde{\omega}, \, \upsilon)
$
and
$
\bR^6 \cong \bR^3 \times \bR^3 \ni u
=
(\omega, \, \upsilon)
$.

Given
a configuration
$q = (R, \, r) \in G$
and
a material velocity
$\dot{q} = (\dot{R}, \, \dot{r}) \in \mathrm{T}_q G$,
the corresponding
\textbf{left-trivialized velocity}
$
\tilde{u} =
(\tilde{\omega}, \, \upsilon)
\in \mathfrak{g}
$
is specifically given by
\begin{equation}
  \begin{aligned}
    G = \mathrm{SO}(3) \times \mathbb{R}^3 \colon \quad&
    (\tilde{\omega}, \, \upsilon)
    \: = \:
    \big(
      R^\textrm{T}\dot{R}, \,
      \dot{r}
    \bigr)
    \\
    G = \mathrm{SO}(3) \ltimes \mathbb{R}^3 \colon \quad&
    (\tilde{\omega}, \, \upsilon)
    \: = \:
    \bigl(
      R^\textrm{T} \dot{R}, \,
      R^\mathrm{T} \dot{r}
    \bigr)
    \,.
  \end{aligned}
  \label{eq:left_trivialization_specialized}
\end{equation}

The different group structures
on the configuration space
lead to different velocity representations.
Given a rigid body motion
$
q_{IB}(t) =
\big( R_{IB}(t), \, {}_Ir_{IB}(t) \big)
$,
the direct product implies that
the left-trivialized translational velocity
is expressed in the basis $I$
of the inertial reference frame, i.e.~%
$\upsilon = {}_I\dot{r}_{IB}$,
while
the semidirect product implies that
it is expressed in the basis $B$
of the body-fixed frame, i.e.~%
$
\upsilon =
R_{BI} \, {}_I\dot{r}_{IB}
$.

As used in~\cref{ssec:pkc},
the map
$\mathrm{T}_e L_q \colon \mathfrak{g} \to \mathrm{T}_q G$
sends a velocity $\tilde{u}$
to its corresponding material velocity
$\mathrm{T}_e L_q(\tilde{u})$.
Moreover,
the dual map
$\mathrm{T}^*_e L_q \colon \mathrm{T}_q^*G \to \mathfrak{g}^*$
sends a force given in the material description
to the equivalent left-trivialized representation.
%
%
The defining property
$
\langle \mathrm{T}_e^* L_q(f_q) \mid \tilde{u} \rangle =
\langle f_q \mid \mathrm{T}_e L_q(\tilde{u}) \rangle
$
for any velocity
$\tilde{u} \in \mathrm{T}_e G$
and any material force covector
$f_q \in \mathrm{T}_q^* G$
hence requires power invariance,
i.e.~force times velocity
is equal in both descriptions.
Based on this,
we have
the left-trivialization
$
\mathrm{T} G \oplus \mathrm{T}^* G
\cong
G \times \mathfrak{g} \times \mathfrak{g}^*
$,
which sends
$
\bigl( q, \, v_q, \, p_q \bigr) \mapsto
\bigl( q, \, \mathrm{T}_q L_{q^{-1}}(v_q), \, \mathrm{T}_e^* L_q(p_q) \bigr)
$.
The symbol $p$ is used here,
as $\mathfrak{g}^*$ is not only the space of
forces, but also the space of momenta,
given in the left-trivialized representation.

\subsection{Adjoint actions}%

The so-called adjoint actions
of the Lie group $G$
reflect the non-commutativity
of its composition operation.
Specifically,
we need
the adjoint action of $G$ on $\mathfrak{g}$
to transform velocities
between different body-fixed frames.
Furthermore,
the adjoint `action' of $\mathfrak{g}$ on itself
gives the so-called Lie bracket,
which makes $\mathfrak{g}$ a Lie algebra.

Both actions are obtained by differentiation
of the conjugation map.
For any $\bar{q} \in G$,
conjugation by
$q \in G$
is defined as
$
\AD_{q}(\bar{q}) =
q \, \bar{q} \, q^{-1}
$.
Equivalently, we have
$
\AD_{q} =
L_q \circ R_{q^{-1}} =
R_{q^{-1}} \circ L_q
$.
%

For a given $q \in G$,
the adjoint action of $G$ on $\mathfrak{g}$
is defined by
\begin{equation*}
  \begin{split}
    \Ad_q \colon
    \mathfrak{g} &\to \mathfrak{g}
    \\
    \tilde{u} &\mapsto \mathrm{T}_e \AD_q(\tilde{u})
    \,.
  \end{split}
\end{equation*}
Specifically, for
$q = (R, \, r) \in G$
and
$
\tilde{u} = (\tilde{\omega}, \, \upsilon)
\in \mathfrak{g}
$,
we have
\begin{equation}
  \begin{aligned}
    G = \mathrm{SO}(3) \times \mathbb{R}^3 \colon \quad&
    \Ad_q(\tilde{u})
    \: = \:
    (R \, \tilde{\omega} \, R^\textrm{T}, \, \upsilon)
    \\
    G = \mathrm{SO}(3) \ltimes \mathbb{R}^3 \colon \quad&
    \Ad_q(\tilde{u})
    \: = \:
    (
      R \, \tilde{\omega} \, R^\textrm{T}, \,
      R \, \upsilon - R \, \tilde{\omega} \, R^\mathrm{T} \, r
    )
    \,.
  \end{aligned}
\end{equation}
Based on
the identification
$
\mathfrak{g}
\cong
\bR^3 \times \bR^3
\ni
u = (\omega, \, \upsilon)
$
and using
the fact
$R \, \tilde{\omega} \, R^\mathrm{T} = \widetilde{R \, \omega}$,
we also write
$
\Ad_q(u) =
\bigl( R \, \omega, \, \upsilon \bigr)
$
in the former
and
$
\Ad_q(u) =
\bigl(
  R \, \omega, \,
  R \, \upsilon - (R \, \omega) \times r
\bigr)
$
in the latter case.

For a given $\tilde{u}_2 \in \mathfrak{g}$,
the adjoint action of $\mathfrak{g}$ on itself
is defined by differentiating
$\Ad_{(\cdot)}(\tilde{u}_2) \colon G \to \mathfrak{g}$,
i.e.~%
$
\ad_{\tilde{u}_1}(\tilde{u}_2) =
(\mathrm{T}_e ( \Ad_{(\cdot)}(\tilde{u}) ))(\tilde{u}_1)
$.
For
$
\tilde{u}_1 = (\tilde{\omega}_1, \, \upsilon_1)
\in \mathfrak{g}
$
and
$
\tilde{u}_2 = (\tilde{\omega}_2, \, \upsilon_2)
\in \mathfrak{g}
$,
we specifically have
\begin{equation*}
  \begin{aligned}
    G = \mathrm{SO}(3) \times \mathbb{R}^3 \colon \quad&
    \ad_{\tilde{u}_1}(\tilde{u}_2)
    \: = \:
    (
      \tilde{\omega}_1 \, \tilde{\omega}_2 - \tilde{\omega}_2 \, \tilde{\omega}_1, \,
      0
    )
    \\
    G = \mathrm{SO}(3) \ltimes \mathbb{R}^3 \colon \quad&
    \ad_{\tilde{u}_1}(\tilde{u}_2)
    \: = \:
    (
      \tilde{\omega}_1 \, \tilde{\omega}_2 - \tilde{\omega}_2 \, \tilde{\omega}_1, \,
      \tilde{\omega}_1 \, \upsilon_2 - \tilde{\omega}_2 \, \upsilon_1
    )
    \,.
  \end{aligned}
\end{equation*}
Based on
the identification
$
\mathfrak{g}
\cong
\bR^3 \times \bR^3
$
and using
the fact
$
\tilde{\omega}_1 \, \tilde{\omega}_2 - \tilde{\omega}_2 \, \tilde{\omega}_1 =
\widetilde{\omega_1 \times \omega_2}
$,
we write
$
\ad_{u_1}(u_2) =
(\omega_1 \times \omega_2, \, 0)
$
in the former
and
$
\ad_{u_1}(u_2) =
(
  \omega_1 \times \omega_2, \,
  \omega_1 \times \upsilon_2 - \omega_2 \times \upsilon_1
)
$
in the latter case.
Equipped with
the Lie bracket
$
\comm{\cdot}{\cdot} \colon
\mathfrak{g} \times \mathfrak{g} \to \mathfrak{g}
$
given by
$
\comm{\tilde{u}_1}{\tilde{u}_2} = \ad_{\tilde{u}_1}(\tilde{u}_2)
$,
the vector space $\mathfrak{g}$
is called the Lie algebra associated to $G$.

\subsection{Coadjoint actions}%

As remarked already,
the map
$\Ad_q \colon \mathfrak{g} \to \mathfrak{g}$
is used to transform
a left-trivialized velocity
between two body-fixed frames,
which are related by a relative pose $q$.
In order to then also transform
the corresponding forces,
we need the dual map
$\Ad^*_q \colon \mathfrak{g}^* \to \mathfrak{g}^*$.
Analogous to the identification
$\mathfrak{g} \cong \bR^3 \times \bR^3 \cong \bR^6$,
we also identify
$\mathfrak{g}^* \cong \bR^3 \times \bR^3 \cong \bR^6$
such that
$
\langle \tilde{f} \mid \tilde{u} \rangle =
\langle f \mid u \rangle =
f^\mathrm{T} u
$
for any
$
f = (f_\omega, \, f_\upsilon)
\in \bR^3 \times \bR^3 \cong \mathfrak{g}^*
$
and
$
u = (\omega, \, \upsilon)
\in \bR^3 \times \bR^3 \cong \mathfrak{g}
$.
We then specifically have
\begin{equation}
  \begin{aligned}
    G = \mathrm{SO}(3) \times \mathbb{R}^3 \colon \quad&
    \Ad^*_q(f)
    \: = \:
    \bigl( R^\mathrm{T} f_\omega, \, f_\upsilon \bigr)
    \\
    G = \mathrm{SO}(3) \ltimes \mathbb{R}^3 \colon \quad&
    \Ad^*_q(f)
    \: = \:
    \bigl(
      R^\mathrm{T} (f_\omega - (r \times f_\upsilon)), \,
      R^\mathrm{T} f_\upsilon
    \bigr)
    \,.
  \end{aligned}
  \label{eq:Ad_star}
\end{equation}

To model the gyroscopic forces
acting on a rigid body
in~\cref{ssec:lp},
we also need
the linear dual of
$\ad_{u} \colon \mathfrak{g} \to \mathfrak{g}$.
%
%
For any
$
p = (p_\omega, \, p_\upsilon)
\in \bR^3 \times \bR^3 \cong \mathfrak{g}^*
$
and
$
u = (\omega, \, \upsilon)
\in \bR^3 \times \bR^3 \cong \mathfrak{g}
$,
we specifically have
\begin{equation}
  \begin{aligned}
    G = \mathrm{SO}(3) \times \mathbb{R}^3 \colon \quad&
    \ad^*_u(p)
    \: = \:
    (p_\omega \times \omega, \, 0)
    \\
    G = \mathrm{SO}(3) \ltimes \mathbb{R}^3 \colon \quad&
    \ad^*_u(p)
    \: = \:
    (
      p_\omega \times \omega + p_\upsilon \times \upsilon, \,
      p_\upsilon \times \omega
    )
    \,.
  \end{aligned}
  \label{eq:ad_star}
\end{equation}

\section{Rigid body model}%
\label{sec:body}


In the first part of this section,
we define
parameters,
configuration and velocity variables
as well as
energy functions
for a rigid body.
In the second part,
we complete the EPHS model of a rigid body.
In the third part,
we show that the resulting evolution equations
agree with
the Lagrange-d'Alembert-Pontryagin principle.

\subsection{Description of the rigid body}%

Let the considered rigid body
$(\cB, \, \rho)$
be characterized by
a set of points
$\cB \subset \cE^3$
that represents its reference configuration
and by
a function
$\rho \colon \cB \rightarrow \bR$
that gives its mass density.

We assume a body-fixed frame $B$
such that the center of mass of the body
coincides with the origin of $B$, i.e.
\begin{equation*}
  \int_{\cB}
  \rho \,
  {}_Br_{BP} \,
  \textrm{d}^3 P
  \: = \:
  0
  \,.
\end{equation*}
Here,
we integrate
over all points $P \in \cB$.
Based on the Euclidian structure,
the integral is computed
separately for each component of ${}_Br_{BP}$.

Let
$q = (R, \, r) = (R_{IB}, \, {}_Ir_{IB}) \in G$
describe the configuration of the body
and
let
$
\dot{q} =
(\dot{R}, \, \dot{r})
= (\dot{R}_{IB}, \, {}_I\dot{r}_{IB})
\in \mathrm{T}_q G
$
be its material velocity.
Concerning left-trivialized quantities,
we henceforth simplify notation
by implicitly identifying $\mathfrak{g}$ and $\bR^6$,
i.e.~we omit the tilde symbol,
whenever there is no ambiguity.
For instance,
we allow ourselves to write
$u = \mathrm{T}_q L_{q^{-1}}(\dot{q}) \in \mathfrak{g}$,
for the velocity.
The same applies for
forces and momenta in
$\mathfrak{g}^* \cong {(\bR^6)}^* \cong \bR^6$.

The kinetic energy of the body
in terms of the material velocity
is given by
\begin{equation*}
  \int_{\cB}
    \frac{1}{2} \,
    \rho \,
    \langle \dot{r}_{IP} , \, \dot{r}_{IP} \rangle \,
  \textrm{d}^3 P
  \: = \:
  \int_{\cB}
    \frac{1}{2} \,
    \rho \,
    \bigl(
      {}_I\dot{r}_{IB} + \dot{R}_{IB} \, {}_Br_{BP}
    \bigr)^\mathrm{T}
    \bigl(
      {}_I\dot{r}_{IB} + \dot{R}_{IB} \, {}_Br_{BP}
    \bigr) \,
  \textrm{d}^3 P
  \: \eqqcolon \:
  T_m(\dot{q})
  \,.
\end{equation*}
We note that
a motion of the particle
with reference position $P \in \cB$
is described by a curve on $\cE^3$.
Based on the Euclidian structure,
we have
$\mathrm{T} \cE^3 \cong \cE^3 \times \bE^3$
and hence
the time derivative of $r_{IP}(t) \in \bE^3$
is considered as a vector in
$\mathrm{T}_I \cE^3 \cong \bE^3$.
On the right hand side
of the above equation,
the basis $I$ is chosen
and the curve
is parametrized based on the configuration $q$.
This gives an expression for
the kinetic energy
$T_m \colon \mathrm{T} G \rightarrow \bR$
in terms of the material velocity.
Rewriting this
in terms of the left-trivialized velocity
gives
the expression
\begin{equation*}
  \int_{\cB}
    \frac{1}{2} \,
    \rho \,
    \bigl(
      \upsilon^\mathrm{T} \upsilon
      \, + \,
      {}_Br_{BP}^\mathrm{T} \, \tilde{\omega}^\mathrm{T}
      \tilde{\omega} \, {}_Br_{BP}
    \bigr) \,
  \textrm{d}^3 P
  \: = \:
  \int_{\cB}
    \frac{1}{2} \,
    \rho \,
    \upsilon^\mathrm{T} \upsilon \,
  \textrm{d}^3 P
  \, + \,
  \int_{\cB}
    \frac{1}{2} \,
    \rho \,
    \omega^\mathrm{T} {}_B\tilde{r}_{BP}^\mathrm{T} \,  \,
    {}_B\tilde{r}_{BP} \, \omega \,
  \textrm{d}^3 P
\end{equation*}
for both group structures alike,
where we use
bilinearity and symmetry of the inner product
as well as~\cref{eq:left_trivialization_specialized}.
The mixed terms vanish since
the center of mass coincides with the origin of $B$.
By factoring
$
u = (\omega, \, \upsilon)
$
out of the integral,
we obtain the
kinetic energy function
$T \colon \mathfrak{g} \rightarrow \bR$
given by
\begin{equation}
  T(u)
  \: = \:
  \frac{1}{2} \,
  \omega^\mathrm{T} \, J \, \omega
  \, + \,
  \frac{1}{2} \, m \,
  \upsilon^\mathrm{T} \, \upsilon
  \label{eq:kinetic_energy}
\end{equation}
with
the total mass $m \in \bR$
and
the moment of inertia tensor
$J \colon \mathfrak{g} \to \mathfrak{g}^*$
given by
\begin{equation*}
  m
  \: = \:
  \int_{\cB} \rho \, \textrm{d}^3 P
  \quad \text{and} \quad
  J
  \: = \:
  \int_{\cB}
  \rho \,
  {}_B\tilde{r}_{BP}^\mathrm{T} \, {}_B\tilde{r}_{BP} \,
  \textrm{d}^3 P
  \,.
\end{equation*}
Because
$\omega$
is expressed in the body frame,
it allows us to compute the kinetic energy with minimal effort,
especially if the basis $B$ is chosen such that
$J$ is diagonal.

While one may choose any function
$V \colon G \to \bR$
that describes
the potential forces acting on the body,
we here assume that
the potential energy function
is given by
\begin{equation}
  V(q)
  \: = \:
  m \, g(r)
  \,,
  \label{eq:potential_energy}
\end{equation}
where
$m$ is again the mass of the body
and
$g \in (\bR^3)^* \cong \bR^3$ is
the gravity force covector
expressed in the dual basis of $I$.
If the basis $I$ is aligned with
the direction of the gravitational force
then $g$ simply picks out
the respective component of $r$.

\subsection{EPHS model of the rigid body}%

We first define
the primitive systems
filling the four inner boxes of
the pattern shown in~\cref{fig:body}
and then
we state
the equations resulting from the composite system.

\subsubsection{Storage of kinetic energy}

The storage component
$(I_\text{ke}, \, E_\text{ke})$
filling the box $\mathtt{ke}$
is defined by
its interface
$
I_\text{ke} =
(\{ \mathtt{p} \}, \, \tau_\text{ke})
$
with
$
\tau_\text{ke}(\mathtt{p}) =
((\mathfrak{g}^*, \, \mathtt{momentum}), \, \mathsf{p})
$
and
its energy function
$
E_\text{ke} \colon \cX_{I_\text{ke}} \rightarrow \bR
$
given by
\begin{equation*}
  E_\text{ke}(p)
  \: = \:
  \frac{1}{2} \,
  p_\omega^\mathrm{T} \, J^{-1} \, p_\omega
  \, + \,
  \frac{1}{2 \, m}
  p_\upsilon^\mathrm{T} \, p_\upsilon
\end{equation*}
for any momentum
$
p = (p_\omega, \, p_\upsilon)
\in \mathfrak{g}^*
$
expressed in the body-fixed frame.
$E_\text{ke}$ is
related to~\cref{eq:kinetic_energy}
by  Legendre transformation, i.e.~%
$
p = (p_\omega, \, p_\upsilon) =
(J \, \omega, \, m \, \upsilon)
$.
The semantics of
$(I_\text{ke}, \, E_\text{ke})$
is then given by
\begin{equation}
  \begin{split}
    \mathtt{ke.p.x}
    \: &= \:
    p
    \: = \:
    (p_\omega, \, p_\upsilon)
    \, \in \, \mathfrak{g}^*
    \\
    \mathtt{ke.p.f}
    \: &= \:
    \dot{p}
    \: = \:
    (\dot{p}_\omega, \, \dot{p}_\upsilon)
    \, \in \, \mathfrak{g}^*
    \\
    \mathtt{ke.p.e}
    \: &= \:
    \dd E_\text{ke}(p)
    \: = \:
    (J^{-1} \, p_\omega, \, m^{-1} \, p_\upsilon)
    \: = \:
    (\omega, \, \upsilon)
    \, \in \, \mathfrak{g}
    \,.
  \end{split}
  \label{eq:body_ke}
\end{equation}

\subsubsection{Gyroscopic effects}
\label{ssec:lp}

The reversible component
$(I_\text{lp}, \, \cD_\text{lp})$
filling the box $\mathtt{lp}$
is defined by
its interface
$
I_\text{lp} =
(\{ \mathtt{p} \}, \, \tau_\text{lp})
$
with
$
\tau_\text{lp}(\mathtt{p}) =
((\mathfrak{g}^*, \, \mathtt{momentum}), \, \mathsf{p})
$
and
its Dirac structure
$\cD_\text{lp}$
given by
\begin{equation}
  \begin{bmatrix}
    \mathtt{lp.p.f}
  \end{bmatrix}
  \: = \:
  \begin{bmatrix}
    -\mathrm{ad}^*_{(\cdot)}(\mathtt{lp.p.x})
  \end{bmatrix}
  \,
  \begin{bmatrix}
    \mathtt{lp.p.e}
  \end{bmatrix}
  \,,
  \label{eq:body_lp}
\end{equation}
with
$\ad^*_{u} \colon \mathfrak{g}^* \to \mathfrak{g}^*$
defined by~\cref{eq:ad_star}.
The net power
at the reversible component
\begin{equation*}
  \langle \mathtt{p.e} \mid \mathtt{p.f} \rangle
  \: = \:
  \langle \mathtt{p.e} \mid -\mathrm{ad}^*_{\mathtt{p.e}}(\mathtt{p.x}) \rangle \\
  \: = \:
  -\langle \mathrm{ad}_{\mathtt{p.e}}(\mathtt{p.e}) \mid \mathtt{p.x} \rangle \\
\end{equation*}
is zero,
since the Lie bracket is antisymmetric
hence
$
\mathrm{ad}_{\mathtt{p.e}}(\mathtt{p.e})
=
\comm{\mathtt{p.e}}{\mathtt{p.e}}
= 0
$.
We note that
the Dirac structure $\cD_\text{lp}$
is 
the Lie-Poisson structure
obtained by symmetry reduction of
the canonical Poisson structure
on $\mathrm{T}^* G$,
see~\cite{1999MarsdenRatiu,1984MarsdenRatiuWeinstein2}.

\subsubsection{Storage of potential energy}

The storage component
$(I_\text{pe}, \, E_\text{pe})$
filling the box $\mathtt{pe}$
is defined by its interface
$
I_\text{pe} =
(\{ \mathtt{q} \}, \, \tau_\text{pe})
$
with
$
\tau_\text{pe}(\mathtt{q}) =
((G, \, \mathtt{pose}), \, \mathsf{p})
$
and
its energy function
$E_\text{pe} = V$,
see~\cref{eq:potential_energy}.
In the absence of potential forces,
we simply let
$E_\mathtt{pe}(q) = 0$.
The semantics of the storage component
is given by
\begin{equation}
  \begin{split}
    \mathtt{q.x}
    \: &= \:
    q
    \: = \:
    (R, \, r)
    \, \in \, G
    \\
    \mathtt{q.f}
    \: &= \:
    \dot{q}
    \: = \:
    (\dot{R}, \, \dot{r})
    \, \in \, \mathrm{T}_q G
    \\
    \mathtt{q.e}
    \: &= \:
    \dd E_\mathtt{pe}(q)
    \: = \:
    \left(
      0, \,
      m \, g
    \right)
    \, \in \, \mathrm{T}^*_q G
  \end{split}
  \label{eq:body_pe}
\end{equation}

\subsubsection{Potential-kinetic coupling}
\label{ssec:pkc}

The reversible component
$(I_\text{pkc}, \cD_\text{pkc})$
is defined by
its interface
$
I_\text{pkc} =
(\{ \mathtt{q}, \, \mathtt{p} \}, \, \tau_\text{pkc})
$
with
$
\tau_\text{pkc}(\mathtt{q}) =
((G, \, \mathtt{pose}), \, \mathsf{p})
$,
$
\tau_\text{pkc}(\mathtt{p}) =
((\mathfrak{g}^*, \, \mathtt{momentum}), \, \mathsf{p})
$
and
its Dirac structure
$\cD_\text{pkc}$
given by
\begin{equation}
  \begin{bmatrix}
    \mathtt{pkc.q.f} \\
    \mathtt{pkc.p.f}
  \end{bmatrix}
  \: = \:
  \begin{bmatrix}
    0 & -\mathrm{T}_e L_q \\
    \mathrm{T}^*_e L_q & 0
  \end{bmatrix}
  \,
  \begin{bmatrix}
    \mathtt{pkc.q.e} \\
    \mathtt{pkc.p.e}
  \end{bmatrix}
  \,,
  \label{eq:body_pkc}
\end{equation}
where $q = \mathtt{pkc.q.x}$.
On the one hand,
this turns
a left-trivialized velocity
$\mathtt{p.e}$
into the corresponding material velocity
$\mathtt{q.f}$
and,
on the other hand,
it turns
a gravity force $\mathtt{q.e}$
given in the material description
into the corresponding
left-trivialized force $\mathtt{p.f}$,
see~\cref{ssec:left_trivialization}.

\subsubsection{Interconnected body model}

By eliminating interface variables,
\cref{eq:body_ke,eq:body_lp,eq:body_pe,eq:body_pkc}
combined with those for
the interconnection pattern shown in~\cref{fig:body}
can be reduced to
\begin{equation}
  \begin{split}
    \dot{q}
    \: &= \:
    \mathrm{T}_e L_q(u)
    \\
    \dot{p}
    \: &= \:
    \mathrm{ad}^*_u(p)
    - \mathrm{T}^*_e L_q(f_q)
    + \mathtt{p.f}
    \\
    \mathtt{p.e}
    \: &= \:
    u
    \,,
  \end{split}
  \label{eq:body}
\end{equation}
where
$u = \dd E_\text{ke}(p)$
and
$f_q = \dd E_\text{pe}(q)$.

\subsection{Variational modeling of the rigid body}%

For modeling a rigid body
as a subsystem of a multibody system,
we use
the Lagrange-d'Alembert-Pontryagin principle
as developed for interconnected Lagrangian systems
in~\cite{2014JacobsYoshimura}.
We here consider the left-trivialized version.
Based on the identification
$\mathrm{T} G \cong G \times \mathfrak{g}$,
let the left-trivialized Lagrangian function
$L \colon G \times \mathfrak{g} \to \bR$
be given by
$L(q, \, u) = T(u) - V(q)$
with the kinetic energy $T$
and the potential energy $V$
defined by~\cref{eq:kinetic_energy,eq:potential_energy}.
For some fixed time interval
$\bI = [t_0, \, t_1] \subset \bR$
and based on the identification
$
\mathrm{T} G \oplus \mathrm{T}^* G \cong
G \times \mathfrak{g} \times \mathfrak{g}^*
$,
we define
the space of smooth curves
\begin{equation*}
  \cC
  \: = \:
  \Bigl\{
    \bigl( q, \, u, \, p \bigr) \, \colon \bI \to
    G \times \mathfrak{g} \times \mathfrak{g}^*
    \, \mid \,
    q(t_0) = q_0
    \, , \,
    q(t_1) = q_1
  \Bigr\}
  \,,
\end{equation*}
where $q_0, \, q_1 \in G$
are some fixed endpoints of the curve $q$.
In some technical sense,
$\cC$ is an infinite-dimensional smooth manifold.
The left-trivialized Hamilton-Pontryagin action
$A \colon \cC \to \bR$
is then defined by
\begin{equation*}
  A(q, \, u, \, p)
  \: = \:
  \int_{t_0}^{t_1} \left(
    L(q, \, u)
    \, + \,
    \langle p \mid \mathrm{T}_q L_{q^{-1}}(\dot{q}) - u \rangle
  \right) \dd t
  \,.
\end{equation*}
Let
$F \in \mathfrak{g}^*$
denote the force
acting on the considered body
due to the tearing / interconnection of the multibody system.
Although we omit the arguments,
$F$ hence is a function of
the configuration and velocity variables of
the interconnected multibody system.
The Lagrange-d'Alembert-Pontryagin principle
requires that
\begin{equation}
  \langle
    \dd A(q, \, u, \, p)
    \mid
    (\delta q, \, \delta u, \, \delta p)
  \rangle
  \, + \,
  \int_{t_0}^{t_1}
  \langle F \mid \eta \rangle \,
  \dd t
  \: = \:
  0
  \label{eq:variational_condition_body}
\end{equation}
for all variations
$
(\delta q, \, \delta u, \, \delta p)
\in \mathrm{T} \cC
$
with extra left-trivialized variation
$\eta = \mathrm{T}_q L_{q^{-1}}(\delta q)$.
Due to the fixed endpoints,
we have
$\delta q(t_0) = \delta q(t_1) = 0$
and hence also
$\eta(t_0) = \eta(t_1) = 0$.

The variational condition~%
\cref{eq:variational_condition_body}
can be written as
\begin{equation*}
  \begin{split}
    \int_{t_0}^{t_1} \Bigl(
      &\langle \dd T(u) \mid \delta u \rangle
      \, + \,
      \langle -\dd V(q) \mid \delta q \rangle
      \, + \,
      \\
      &\langle \delta p \mid \mathrm{T}_q L_{q^{-1}}(\dot{q}) - u \rangle
      \, + \,
      \langle p \mid \delta \bigl( \mathrm{T}_q L_{q^{-1}}(\dot{q}) \bigr) \rangle
      \, + \,
      \langle -p \mid \delta u \rangle
      \, + \,
      \langle F \mid \eta \rangle \,
    \Bigr) \,
    \dd t
    \: = \:
    0
    \,.
  \end{split}
\end{equation*}
Using the identity
$
\delta \bigl( \mathrm{T}_q L_{q^{-1}}(\dot{q}) \bigr) =
\dot{\eta} + \ad_u(\eta)
$,
see~\cite{1991MarsdenRatiuRaugel,1996BlochKrishnaprasadMarsdenRatiu},
and
partial integration on the term containing
$\dot{\eta}$,
this can be further transformed into
\begin{equation*}
  \begin{split}
    \int_{t_0}^{t_1} \Bigl(
      &\langle \dd T(u) \mid \delta u \rangle
      \, + \,
      \langle -\mathrm{T}_e^* L_q(\dd V(q)) \mid \eta \rangle
      \, + \,
      \\
      &\langle \delta p \mid \mathrm{T}_q L_{q^{-1}}(\dot{q}) - u \rangle
      \, + \,
      \langle -\dot{p} + \ad_u^*(p) \mid \eta \rangle
      \, + \,
      \langle -p \mid \delta u \rangle
      \, + \,
      \langle F \mid \eta \rangle \,
    \Bigr) \,
    \dd t
    \: = \:
    0
    \,.
  \end{split}
\end{equation*}
From this, it follows that
\begin{equation*}
  \begin{split}
    \dot{q}
    \: &= \:
    \mathrm{T}_e L_q(u)
    \\
    p
    \: &= \:
    \dd T(u)
    \\
    \dot{p}
    \: &= \:
    \ad_u^*(p)
    -\mathrm{T}_e^* L_q \bigl( \dd V(q) \bigr)
    + F
    \,.
  \end{split}
\end{equation*}
It can be easily checked that
this is equivalent to~\cref{eq:body}
with $\mathtt{p.f} = F$.

\section{Joint model}%
\label{sec:joint}

We proceed as for the rigid body.
First,
we describe the joint
and
then
we complete its EPHS model.
Finally,
we show that
a variational modeling approach
yields the same evolution equations.

\subsection{Description of the joint}%

The considered joint connects two bodies.
The body-fixed frame $B$
of the first body
and its corresponding configuration
$
q
$
are here denoted by
$B_1$
and
$
q_1 = (R_1, \, r_1) =
(R_{IB_1}, \, {}_Ir_{IB_1}) \in G
$.
Analogously,
the frame and the configuration
of the second body are denoted by
$B_2$
and
$
q_2 = (R_2, \, r_2) =
(R_{IB_2}, \, {}_Ir_{IB_2}) \in G
$.

For each body, we define a second
body-fixed frame,
whose origin is located at
the joint force application point.
For the first body,
the joint frame
and
the corresponding configuration
are denoted by
$C_1$
and
$
q_{j1} =
(R_{IC_1}, \, {}_Ir_{IC_1}) \in G
$.
For the second body,
we analogously have
$C_2$
and
$
q_{j2} =
(R_{IC_2}, \, {}_Ir_{IC_2}) \in G
$.

We want to describe the fixed offset between
$B_1$ and $C_1$
by a relative pose
$
o_1
\in G
$
satisfying
$q_{j1} = q_1 \, o_1$.
As discussed in~\cref{ssec:composition_interpretation},
this implies
\begin{equation*}
  \begin{aligned}
    G = \mathrm{SO}(3) \times \mathbb{R}^3 \colon \quad&
    o_1
    \: = \:
    (R_{B_1C_1}, \, {}_Ir_{B_1C_1})
    \\
    G = \mathrm{SO}(3) \ltimes \mathbb{R}^3 \colon \quad&
    o_1
    \: = \:
    (R_{B_1C_1}, \, {}_{B_1}r_{B_1C_1})
    \,.
  \end{aligned}
\end{equation*}
The parameter $o_1$,
which describes
where the joint is attached to the body,
is constant only if we choose the semidirect product.
For the direct product,
the description of the fixed offset
depends on the configuration of the body $q_1$,
making the parameter a function
$
o_1(q_1) =
(R_{B_1C_1}, \, R_{IB_1} \, {}_{B_1}r_{B_1C_1})
$.
The offset $o_2$ between $B_2$ and $C_2$
is defined analogously.

The model assumes that
the set of all relative poses of the two bodies,
which are permitted by the joint,
form a Lie subgroup $\bar{G}$ of $G$.
In other words,
the joint model implements
a lower kinematic pair, i.e.~%
either a spherical, planar, cylindrical, revolute, prismatic, or screw joint.
We denote the inclusion functions
for
the subgroup $\bar{G}$
and
its associated Lie algebra $\bar{\mathfrak{g}}$
by
\begin{equation*}
  \begin{alignedat}{2}
    \mathrm{I} &\colon \bar{G} &&\hookrightarrow G\\
    \mathrm{i} &\colon \bar{\mathfrak{g}} &&\hookrightarrow \mathfrak{g}
    \,,
  \end{alignedat}
\end{equation*}
where
$\mathrm{i} = \mathrm{T}_e \mathrm{I}$.
The relative pose $q_r \in \bar{G}$
then satisfies the holonomic constraint
\begin{equation*}
  q_{j2}
  \: = \:
  q_{j1} \, \mathrm{I}(q_r)
  \,.
\end{equation*}
Analogously to the fixed offsets $o_1$ and $o_2$,
this implies
\begin{equation*}
  \begin{aligned}
    G = \mathrm{SO}(3) \times \mathbb{R}^3 \colon \quad&
    \mathrm{I}(q_r)
    \: = \:
    (R_{C_1C_2}, \, {}_Ir_{C_1C_2})
    \\
    G = \mathrm{SO}(3) \ltimes \mathbb{R}^3 \colon \quad&
    \mathrm{I}(q_r)
    \: = \:
    (R_{C_1C_2}, \, {}_{C_1}r_{C_1C_2})
    \,.
  \end{aligned}
\end{equation*}
In the case of the direct product,
the relative pose of the two bodies
is not described independently of their absolute pose
and thus
it does not directly lie in a subgroup.
The relative pose thus
needs to be defined as
a function
not only of $q_r$,
but also of $q_{j1}$.

We choose to
henceforth restrict our attention to
the simpler case
$G = \mathrm{SO}(3) \ltimes \mathbb{R}^3$.
The motivation for this is twofold.
First,
presenting the models also for the direct product
would make the paper
significantly longer
and less easy to follow,
as we would have to
define different interconnection patters
with extra state ports that share
the additionally required configuration variables.
Also,
various components would have to be defined differently
for the two cases.
Second,
it is argued
in~\cite{2014SonnevilleBruls,2015Sonneville}
that using a truly relative description for the joints
also leads to
important advantages for numerical simulation.

Summarizing the above
for the henceforth considered case
$G = \mathrm{SO}(3) \ltimes \mathbb{R}^3$,
the involved configuration variables
are related by
the three constraints
\begin{equation*}
  \begin{split}
    q_{j1}
    \: &= \:
    q_1 \, o_1
    \\
    q_{j2}
    \: &= \:
    q_2 \, o_2
    \\
    q_{j2}
    \: &= \:
    q_{j1} \, \mathrm{I}(q_r)
    \,,
  \end{split}
\end{equation*}
where the constant relative poses
$o_1, \, o_2 \in G$
are parameters
and $q_r \in \bar{G}$
is regarded as
the configuration of the joint itself.
We now express these constraints
on the velocity level.
Differentiating the first constraint yields
$\dot{q}_{j1} = \mathrm{T}R_{o_1}(\dot{q}_1)$,
where $R_{o_1}$ denotes right translation by $o_1$.
Rewriting this in terms
the corresponding left-trivialized velocities
gives
$
\mathrm{T} L_{q_{j1}}(u_{j1})
\: = \:
( \mathrm{T}R_{o_1} \circ \mathrm{T} L_{q_1} )(u_1)
$.
Solving for $u_{j1}$
and
using commutativity of left and right translations
gives
$
u_{j1} = \Ad_{o_1^{-1}}(u_1)
$.
Analogously,
differentiating the third constraint yields
$
\dot{q}_{j2} =
\mathrm{T}R_{\mathrm{I}(q_r)}(\dot{q}_{j1}) +
(\mathrm{T}L_{q_{j1}} \circ \mathrm{T}I)(\dot{q}_r)
$.
Using
$
\mathrm{T}\mathrm{I}(\dot{q}_r) =
\mathrm{T}L_{\mathrm{I}(q_r)}(i(u_r))
$,
we can write this as
$
u_{j2} =
\Ad_{\mathrm{I}(q_r^{-1})}(u_{j1}) + i(u_r)
$.
Hence,
the three constraints
in terms of left-trivialized velocities
are given by
\begin{equation}
  \begin{split}
    u_{j1}
    \: &= \:
    \Ad_{o_1^{-1}}(u_1)
    \\
    u_{j2}
    \: &= \:
    \Ad_{o_2^{-1}}(u_2)
    \\
    0
    \: &= \:
    \Ad_{\mathrm{I}(q_r^{-1})}(u_{j1}) - u_{j2} + i(u_r)
    \,.
  \end{split}
  \label{eq:joint_constraints}
\end{equation}

The model
may include potential forces
that depend on
the relative configuration of the two bodies.
If no such forces are present,
we simply let
the potential energy
$V_r \colon \bar{G} \to \bR$
be given by
$V_r(q_r) = 0$.

Finally,
the joint friction
is modeled by
a non-negative definite $2$-covariant tensor $\mu$.
Seen as a linear map
$\mu^\flat : \bar{\mathfrak{g}} \to \bar{\mathfrak{g}}^*$,
it determines
the friction force
$F^\text{fr}(u_r) = -\mu^\flat(u_r)$.
Hence,
the dissipated power is given by
$
\langle \mu^\flat(u_r) \mid u_r \rangle =
\mu(u_r, \, u_r) \geq 0
$.
For a non-linear
and possibly temperature-dependent
friction model
we would define
$\mu$ as a tensor-valued function.

\subsection{EPHS model of the joint}%

We first define the primitive systems
filling the seven inner boxes
of the pattern shown in~\cref{fig:joint}
and then we state the equations
resulting from the composite system.

\subsubsection{Relative pose and storage of potential energy}

The storage component
$(I_\text{pe}, \, E_\text{pe})$
filling the box $\mathtt{pe}$
is defined by its interface
$
I_\text{pe} =
(\{ \mathtt{q_r} \}, \, \tau_\text{pe})
$
with
$
\tau_\text{pe}(\mathtt{q_r}) =
((\bar{G}, \, \mathtt{relative\_pose}), \, \mathsf{p})
$
and
its energy function
$E_\text{pe} = V_r$.
The semantics of the storage component
is thus given by
\begin{equation}
  \begin{split}
    \mathtt{pe.q_r.x}
    \: &= \:
    q_r
    \: \in \: \bar{G}
    \\
    \mathtt{pe.q_r.f}
    \: &= \:
    \dot{q}_r
    \: \in \: \mathrm{T}_{q_r} \bar{G}
    \\
    \mathtt{pe.q_r.e}
    \: &= \:
    \dd V_r(q_r)
    \: \in \: \mathrm{T}^*_{q_r} \bar{G}
    \,.
  \end{split}
  \label{eq:joint_pe}
\end{equation}

\subsubsection{Potential-kinetic coupling}

The reversible component
$(I_\text{pkc}, \, \cD_\text{pkc})$
filling the box $\mathtt{pkc}$
is defined by
its interface
$
I_\text{pkc} =
(\{ \mathtt{q_r}, \, \mathtt{p_r} \}, \, \tau_\text{pkc})
$
with
$
\tau_\text{pkc}(\mathtt{q_r}) =
((\bar{G}, \, \mathtt{relative\_pose}), \, \mathsf{p})
$,
$
\tau_\text{pkc}(\mathtt{p_r}) =
((\bar{\mathfrak{g}}^*, \, \mathtt{momentum}), \, \mathsf{p})
$
and
its Dirac structure
$\cD_\text{pkc}$
given by
\begin{equation}
  \begin{bmatrix}
    \mathtt{pkc.q_r.f} \\
    \mathtt{pkc.p_r.f}
  \end{bmatrix}
  \: = \:
  \begin{bmatrix}
    0 & -\mathrm{T}_e L_{q_r} \\
    \mathrm{T}^*_e L_{q_r} & 0
  \end{bmatrix}
  \,
  \begin{bmatrix}
    \mathtt{pkc.q_r.e} \\
    \mathtt{pkc.p_r.e}
  \end{bmatrix}
  \,,
  \label{eq:joint_pkc}
\end{equation}
where $q_r = \mathtt{pkc.q_r.x}$.

\subsubsection{Offsets}

The reversible component
$(I_{\text{o}_1}, \, \cD_{\text{o}_1})$
filling the box $\mathtt{o_1}$
is defined by
its interface
$
I_{\text{o}_1} =
(\{ \mathtt{p_1}, \, \mathtt{p_{j1}} \}, \, \tau_{\text{o}_1})
$
with
$
\tau_{\text{o}_1}(\mathtt{p_1}) =
\tau_{\text{o}_1}(\mathtt{p_{j1}}) =
((\mathfrak{g}^*, \, \mathtt{momentum}), \, \mathsf{p})
$
and
its Dirac structure
$\cD_{\text{o}_1}$
given by
\begin{equation}
  \left[
    \begin{array}{c}
      \mathtt{o_1.p_1.f} \\ \hline
      \mathtt{o_1.p_{j1}.e}
    \end{array}
  \right]
  \: = \:
  \left[
    \begin{array}{c | c}
      0 & -\mathrm{Ad}_{o_1^{-1}}^* \\ \hline
      \mathrm{Ad}_{o_1^{-1}} & 0
    \end{array}
  \right]
  \,
  \left[
    \begin{array}{c}
      \mathtt{o_1.p_1.e} \\ \hline
      \mathtt{o_1.p_{j1}.f}
    \end{array}
  \right]
  \,.
  \label{eq:joint_o1}
\end{equation}
This describes
the transformation of
the velocity $\mathtt{p_1.e}$
of the first body
expressed in frame $B_1$
to the same velocity given in the joint frame $C_1$
and
dually
the transformation of
the joint forces $\mathtt{p_{j1}.f}$
expressed in frame $C_1$
to the same forces expressed in $B_1$.

The reversible component
filling the box $\mathtt{o_2}$
is defined analogously
using the offset parameter $o_2$.

\subsubsection{Holonomic constraint}

The reversible component
$(I_\text{hc}, \, \cD_\text{hc})$
filling the box $\mathtt{hc}$
is defined by
its interface
$
I_\text{hc} =
(
  \{ \mathtt{q_r}, \, \mathtt{p_r}, \, \mathtt{p_{j1}}, \, \mathtt{p_{j2}} \}, \,
  \tau_\text{hc}
)
$
with
$
\tau_\text{hc}(\mathtt{q_r}) =
((\bar{G}, \, \mathtt{relative\_pose}), \, \mathsf{s})
$,
$
\tau_\text{hc}(\mathtt{p_r}) =
((\bar{\mathfrak{g}}^*, \, \mathtt{momentum}), \, \mathsf{p})
$,
$
\tau_\text{hc}(\mathtt{p_{j1}}) =
\tau_\text{hc}(\mathtt{p_{j2}}) =
((\mathfrak{g}^*, \, \mathtt{momentum}), \, \mathsf{p})
$
and
its Dirac structure
$\cD_\text{hc}$
given by
\begin{equation}
  \begin{bmatrix}
    f \\
    0
  \end{bmatrix}
  \: = \:
  \begin{bmatrix}
    & C^*(q_r) \\
    -C(q_r) &
  \end{bmatrix}
  \,
  \begin{bmatrix}
    e \\
    \lambda_c
  \end{bmatrix}
  \,,
  \label{eq:joint_hc}
\end{equation}
where
\begin{equation*}
  \begin{split}
    f
    \: &= \:
    \bigl(
      \mathtt{hc.p_{j1}.f}, \, \mathtt{hc.p_{j2}.f}, \, \mathtt{hc.p_r.f}
    \bigr)
    \\
    e
    \: &= \:
    \bigl(
      \mathtt{hc.p_{j1}.e}, \, \mathtt{hc.p_{j2}.e}, \, \mathtt{hc.p_r.e}
    \bigr)
    \\
    q_r
    \: &= \:
    \mathtt{hc.q_r.x}
  \end{split}
\end{equation*}
and
\begin{equation*}
  C(q_r)
  \: = \:
  \begin{bmatrix}
    \mathrm{Ad}_{\mathrm{I}(q_r^{-1})}
    &
    -\mathrm{id}_{\mathfrak{g}}
    &
    \mathrm{i}
  \end{bmatrix}
  \,.
\end{equation*}
The
Lagrange multiplier
$\lambda_c$
hence
represents the joint forces
expressed in frame $C_2$.

\subsubsection{Mechanical friction}

The irreversible component
$(I_\text{mf}, \, \cO_\text{mf})$
filling the box $\mathtt{mf}$
is defined by
its interface
$
I_\text{mf} =
(\{ \mathtt{p_r}, \, \mathtt{s} \}, \, \tau_\text{mf})
$
with
$
\tau_\text{mf}(\mathtt{p_r}) =
((\bar{\mathfrak{g}}^*, \, \mathtt{momentum}), \, \mathsf{p})
$,
$
\tau_\text{mf}(\mathtt{s}) =
((\bR, \, \mathtt{entropy}), \, \mathsf{p})
$
and its Onsager structure $\cO_\text{mf}$
given by
\begin{equation}
    \left[
      \begin{array}{c}
        \mathtt{mf.p_r.f} \\
        \mathtt{mf.s.f}
      \end{array}
    \right]
    \: = \:
    \frac{1}{\color{violet} \theta_0} \,
    \left[
      \begin{array}{cc}
        \theta \, \mu^\flat(\cdot)
        &
        -\mu(u_r, \, \cdot)
        \\
        -\mu(u_r, \, \cdot)
        &
        \frac{(\cdot)}{\theta} \, \mu(u_r, \, u_r)
      \end{array}
    \right]
    \,
    \left[
      \begin{array}{c}
        \mathtt{mf.p_r.e} \\
        \mathtt{mf.s.e}
      \end{array}
    \right]
    \: = \:
    \left[
      \begin{array}{c}
        \mu^\flat(u_r) \\
        -\frac{1}{\theta} \, \mu(u_r, \, u_r)
      \end{array}
    \right]
    \,,
  \label{eq:joint_mf}
\end{equation}
where
$u_r = \mathtt{p_r.e}$
is the relative velocity of the two bodies
expressed in frame $C_1$
and
$\theta = \textcolor{violet}{\theta_0} + \mathtt{s.e}$
is the absolute temperature
at which heat is dissipated.
The exergy destruction rate is given by
\begin{equation*}
  \langle \mathtt{p_r.e} \mid \mathtt{p_r.f} \rangle +
  \langle \mathtt{s.e} \mid \mathtt{s.f} \rangle
  \: = \:
  \textcolor{violet}{\theta_0} \,
  \frac{1}{\theta} \, \mu(u_r, \, u_r)
  \: \geq \: 0
\end{equation*}
and
energy is conserved
since
\begin{equation*}
  \left[
    \begin{array}{cc}
      \theta \, \mu^\flat(\cdot)
      &
      -\mu(u_r, \, \cdot)
      \\
      -\mu(u_r, \, \cdot)
      &
      \frac{(\cdot)}{\theta} \, \mu(u_r, \, u_r)
    \end{array}
  \right]
  \,
  \left[
    \begin{array}{c}
      u_r \\
      \theta
    \end{array}
  \right]
  \: = \:
  \left[
    \begin{array}{c}
      0 \\
      0
    \end{array}
  \right]
  \,.
\end{equation*}

\subsubsection{Environment}

The semantics of
the environment component
filling the box $\mathtt{env}$
is given by
\begin{equation}
  \begin{split}
    \mathtt{env.s.x}
    \: &= \:
    s
    \\
    \mathtt{env.s.f}
    \: &= \:
    \dot{s}
    \\
    \mathtt{env.s.e}
    \: &= \:
    0
    \,.
  \end{split}
  \label{eq:joint_env}
\end{equation}

\subsubsection{Interconnected joint model}

By eliminating interface variables,
\cref{eq:joint_pe,eq:joint_pkc,eq:joint_o1,eq:joint_hc,eq:joint_mf,eq:joint_env}
combined with those for
the interconnection pattern shown in~\cref{fig:joint}
can be reduced to
\begin{equation}
  \begin{split}
    \dot{q}_r
    \: &= \:
    \mathrm{T}_e L_{q_r}(u_r)
    \\
    \dot{s}
    \: &= \:
    \frac{1}{\theta_0} \, \mu(u_r, \, u_r)
    \\
    0
    \: &= \:
    \mathrm{T}_e^* L_{q_r} \bigl( \dd V_r(q_r) \bigr)
    + \mathrm{i}^*(\lambda_c)
    + \mu^\flat(u_r)
    \\
    0
    \: &= \:
    \bigl( \Ad_{\mathrm{I}(q_r^{-1})} \circ \Ad_{o_1^{-1}} \bigr)(\mathtt{p_1.e})
    -\Ad_{o_2^{-1}}(\mathtt{p_2.e})
    +\mathrm{i}(u_r)
    \\
    \mathtt{p_1.f}
    \: &= \:
    \bigl( \Ad_{o_1^{-1}}^* \circ \Ad_{\mathrm{I}(q_r^{-1})}^* \bigr)(\lambda_c)
    \\
    \mathtt{p_2.f}
    \: &= \:
    -\Ad_{o_2^{-1}}^*(\lambda_c)
    \,.
  \end{split}
  \label{eq:joint}
\end{equation}
%

\subsection{Variational modeling of the joint}%

For modeling a joint
as a subsystem of a multibody system,
we again use
the left-trivialized Lagrange-d'Alembert-Pontryagin principle,
this time with
the mechanical constraints
in~\cref{eq:joint_constraints}
as well as
a constraint of thermodynamic type~\cite{2020GayYoshimura}
that describes the irreversible process
of mechanical friction.

To control the plethora of variables,
we choose to eliminate
$q_{j1}$ and $q_{j2}$ upfront.
We hence consider
the configuration space
$
\cQ =
G \times G \times \bar{G}
\ni
q = (q_1, \, q_2, \, q_r)
$.
Regarding the left-trivialized tangent bundle
and velocities,
we have
$
\mathrm{T} \cQ
\cong
\cQ \times U
$
with
$
U =
\mathfrak{g} \times \mathfrak{g} \times \bar{\mathfrak{g}}
\ni u = (u_1, \, u_2, \, u_r)
$.
The Lagrangian
$L \colon \cQ \times U \times \bR \to \bR$
is given by
\begin{equation*}
  L(q, \, u, \, s)
  \: = \:
  -V_r(q_r)
  -U(s)
  \,,
\end{equation*}
where $U(s) = \theta_0 \, s$
is the internal energy of
the isothermal environment.
The left-trivialized velocities,
which are permitted by the joint
when it is in the configuration $q_r$
are given by
$
\Delta_m(q_r) =
\{
  u \in U
  \, \mid \,
    \bigl( \Ad_{\mathrm{I}(q_r^{-1})} \circ \Ad_{o_1^{-1}} \bigr)(u_1)
    -\Ad_{o_2^{-1}}(u_2)
    +\mathrm{i}(u_r)
    = 0
\}
$,
as can be seen by combining
the three constraints in~\cref{eq:joint_constraints}.
Regarding the left-trivialized cotangent bundle
and momenta as well as forces,
we have
$
\mathrm{T}^* \cQ
\cong
\cQ \times U^*
$
with
$
U^* \ni p = (p_1, \, p_2, \, p_r)
$
and
$
U^* \ni f = (f_1, \, f_2, \, f_r)
$.
The admissible constraint forces
do no virtual work
for admissible virtual displacements
and
are consequently given by
\begin{equation*}
  \begin{alignedat}{2}
    \Delta_m^\circ(q_r)
    \: &= \:
    \Bigl\{
      f \in U^*
      \, \mid \,
      &&\forall \eta \in \Delta_m(q_r) :
      \langle f \mid \eta \rangle = 0
    \Bigr\}
    \\
    \: &= \:
    \Bigl\{
      f \in U^*
      \, \mid \,
      &&f_1 =
      \bigl( \Ad_{o_1^{-1}}^* \circ \Ad_{\mathrm{I}(q_r^{-1})}^* \bigr)(\lambda_c)
      \, , \,
      \\
      & &&f_2 =
      -\Ad_{o_2^{-1}}^*(\lambda_c)
      \, , \,
      \\
      & &&f_r =
      \mathrm{i}^*(\lambda_c)
      \, , \,
      \lambda_c \in \mathfrak{g}^*
    \Bigr\}
    \,.
  \end{alignedat}
\end{equation*}
For some fixed time interval
$\bI = [t_0, \, t_1] \subset \bR$,
we define
the space of smooth curves
\begin{equation*}
  \begin{alignedat}{2}
    \cC
    \ : = \:
    \Bigl\{
      \bigl( q, \, u, \, p, \, s, \, u_s, \, p_s \bigr) \, \colon \bI \to
      \cQ \times U \times U^*
      \times
      \bR \times \bR \times \bR^*
      \mid \,
      &q(t_0) = q_0
      \, , \,
      &&q(t_1) = q_1
      \, , \,
      \\
      &s(t_0) = s_0
      \, , \,
      &&s(t_1) = s_1
    \Bigr\}
    \,,
  \end{alignedat}
\end{equation*}
where $q_0, \, q_1 \in \cQ$
and $s_0, \, s_1 \in \bR$
are fixed endpoints
for the curves $q$ and $s$.
The left-trivialized Hamilton-Pontryagin action
$A \colon \cC \to \bR$
is then defined by
\begin{equation*}
  \begin{split}
    A(q, \, u, \, p, \, s, \, u_s, \, p_s)
    =
    \int_{t_0}^{t_1} \Bigl(
      L(q, \, u, \, s)
      \, + \,
      &\langle p_1 \mid \mathrm{T} L_{q_1^{-1}}(\dot{q}_1) - u_1 \rangle
      \, + \,
      \langle p_2 \mid \mathrm{T} L_{q_2^{-1}}(\dot{q}_2) - u_2 \rangle
      \\
      \, + \,
      &\langle p_r \mid \mathrm{T} L_{q_r^{-1}}(\dot{q}_r) - u_r \rangle
      \, + \,
      \langle p_s \mid \dot{s} - u_s \rangle
    \Bigr) \, \dd t
    \,.
  \end{split}
\end{equation*}
Let
$F_{j1}, \, F_{j2} \in \mathfrak{g}^*$
denote the joint forces
acting on the two connected bodies.
Although we omit the arguments,
they are functions of
the configuration and velocity variables of
the interconnected multibody system.
The Lagrange-d'Alembert-Pontryagin principle
requires that
\begin{equation}
  \begin{split}
    \langle
      \dd A(q, \, u, \, p, \, s, \, u_s, \, p_s)
      \mid
      (\delta q, \, \delta u, \, \delta p, \, \delta s, \, \delta u_s, \, \delta p_s)
    \rangle
    \: &+ \:
    \\
    \int_{t_0}^{t_1}
    \Bigl(
      \langle F_{j1} \mid \eta_1 \rangle \,
      \, + \,
      \langle F_{j2} \mid \eta_2 \rangle \,
    \Bigr) \,
    \dd t
    \: &= \:
    0
  \end{split}
  \label{eq:variational_condition_joint}
\end{equation}
for all variations
$
(\delta q, \, \delta u, \, \delta p, \, \delta s, \, \delta u_s, \, \delta p_s)
\in \mathrm{T} \cC
$
with extra left-trivialized variations
$
\eta =
(\eta_1, \, \eta_2, \, \eta_r)
= \bigl(
  \mathrm{T} L_{q_1^{-1}}(\delta q_1), \,
  \mathrm{T} L_{q_2^{-1}}(\delta q_2), \,
  \mathrm{T} L_{q_r^{-1}}(\delta q_r)
\bigr)
$.
Additionally,
the curves $q$ and $u$
are subject to the mechanical constraint
$
u \in \Delta_m(q_r)
$
with corresponding variational constraint
$
\eta \in \Delta_m(q_r)
$.
Moreover,
the curves $s$ and $u$
are subject to the thermodynamic constraint
$
\langle \frac{\partial L}{\partial s}(s) \mid u_s \rangle =
\langle F^\text{fr}(u_r) \mid u_r \rangle
$
with corresponding variational constraint
\begin{equation*}
  \langle \frac{\partial L}{\partial s}(s) \mid \delta s \rangle
  \: = \:
  \langle F^\text{fr}(u_r) \mid \eta_r \rangle
  \,.
\end{equation*}
Finally, we note that
due to the fixed endpoints,
we have
$\delta q(t_0) = \delta q(t_1) = 0$
and hence also
$\eta(t_0) = \eta(t_1) = 0$
as well as
$\delta s(t_0) = \delta s(t_1) = 0$.

The variational condition~\cref{eq:variational_condition_joint}
can be written as
\begin{equation*}
  \begin{split}
    \int_{t_0}^{t_1} \Biggl(
      &\Bigl(
        \langle -\dd V_r(q_r) \mid \delta q_r \rangle
        \, + \,
        \langle \frac{\partial L}{\partial s}(s) \mid \delta s \rangle
      \Bigr)
      \, + \,
      \\
      &\Bigl(
        \langle \delta p_1 \mid \mathrm{T}_{q_1} L_{q_1^{-1}}(\dot{q_1}) - u_1 \rangle
        \, + \,
        \langle p_1 \mid \delta \bigl( \mathrm{T}_{q_1} L_{q_1^{-1}}(\dot{q_r}) \bigr) \rangle
        \, + \,
        \langle -p_1 \mid \delta u_1 \rangle
        \, + \,
        \ldots
      \Bigr)
      \, + \,
      \\
      &\Bigl(
        \langle F_{j1} \mid \eta_1 \rangle
        \, + \,
        \langle F_{j2} \mid \eta_2 \rangle
      \Bigr)
      \, + \,
      \\
      &\Bigl(
        \langle \bigl( \Ad_{o_1^{-1}}^* \circ \Ad_{\mathrm{I}(q_r^{-1})}^* \bigr)(\lambda_c) \mid \eta_1 \rangle
        \, + \,
        \langle -\Ad_{o_2^{-1}}^*(\lambda_c) \mid \eta_2 \rangle
        \, + \,
        \langle \mathrm{i}^*(\lambda_c) \mid \eta_r \rangle
      \Bigr)
    \Biggr) \,
    \dd t
    \: = \:
    0
    \,.
  \end{split}
\end{equation*}
In the first line,
we rewrite the first term in terms of $\eta_r$
and
we replace the second term with
$\langle F^\text{fr}(u_r) \mid \eta_r \rangle$.
In the second line,
the ellipsis represents the three analogous terms.
Since there is no kinetic energy in the joint
and $p_s = 0$ in general,
only the kinematic constraints
$\dot{q}_1 = \mathrm{T}_e L_{q_1}(u_1)$, $\ldots$,
as well as
$\dot{s} = u_s$ remain.
The last line is present since
by definition of $\Delta_m^\circ$,
any admissible constraint force
$f \in \Delta_m^\circ(q_r)$
can be added without changing
the left hand side of the variational condition.
From this,
we again obtain~\cref{eq:joint}
with
$\mathtt{p_1.f} = F_{j1}$
and
$\mathtt{p_2.f} = F_{j2}$.

\section{Basic multibody system}%
\label{sec:mbs}

As stated in~\cref{fig:mbs},
the considered multibody system
comprises
two bodies,
as defined in~\cref{sec:body},
which are connected by a joint,
as defined in~\cref{sec:joint}.

By eliminating interface variables,
\cref{eq:body} for the bodies
and~\cref{eq:joint} for the joint
combined with the equations for
the pattern in~\cref{fig:mbs}
give
\begin{align*}
  \dot{q}_1
  \: &= \:
  + \mathrm{T}_e L_{q_1}(u_1)
  \\
  \dot{p}_1
  \: &= \:
  + \mathrm{ad}^*_{u_1}(p_1)
  - \mathrm{T}^*_e L_{q_1}(f_{q1})
  - (\mathrm{Ad}^*_{o_1^{-1}} \circ \mathrm{Ad}^*_{\mathrm{I}(q_r^{-1})})(\lambda_c)
  \\[1em]
  \dot{q}_2
  \: &= \:
  + \mathrm{T}_e L_{q_2}(u_2)
  \\
  \dot{p}_2
  \: &= \:
  + \mathrm{ad}^*_{u_2}(p_2)
  - \mathrm{T}^*_e L_{q_2}(f_{q2})
  + \mathrm{Ad}^*_{o_2^{-1}}(\lambda_c)
  \\[1em]
  \dot{q}_r
  \: &= \:
  + \mathrm{T}_e L_{q_r}(u_r)
  \\
  0
  \: &= \:
  - \mathrm{T}^*_e L_{q_r}(f_{qr})
  - \mu^\flat(u_r)
  - \mathrm{i}^*(\lambda_c)
  \\[1em]
  0
  \: &= \:
  + (\mathrm{Ad}_{\mathrm{I}(q_r^{-1})} \circ \mathrm{Ad}_{o_1})(u_1)
  - \mathrm{Ad}_{o_2}(u_2)
  + \mathrm{i}(u_r)
  \\[1em]
  \dot{s}
  \: &= \:
  + \frac{1}{\theta_0} \, \mu(u_r, \, u_r)
  \,,
\end{align*}
where
$u_1 = \dd E_{\text{ke},1}(p_1)$,
$u_2 = \dd E_{\text{ke},2}(p_2)$,
$f_{q1} = \dd V_1(q_1)$,
$f_{q2} = \dd V_2(q_2)$ and
$f_{qr} = \dd V_r(q_r)$.

The same equations are obtained by
concatenating the variational formulations
for the bodies as well as the joint
and adding
interconnection constraints
that identify the velocities
$u_1$ and $u_2$
in the body and the joint models.
Concerning the admissible forces,
we then have
$F_1 + F_{j1} = 0$
and
$F_2 + F_{j2} = 0$,
where $F_1$ and $F_2$ denote
the external forces in the two body models.

\section{Discussion}%
\label{sec:discussion}

It is our hope that
the EPHS language
makes it possible to
develop more complex models
based, among others, on the presented building blocks,
without necessarily requiring
a deep understanding of the involved mathematics.
Also beneficial for educational purposes,
it seems that models can be understood
simply by understanding
the physical meaning of the components
and their port variables.

Interesting directions for further research
include
the natural discretization of the presented models.
Naturality here is a condition requiring that
discretization and interconnection commute.
Since joint constraints are enforced
solely on the velocity level,
a numerical drift can be expected in simulations.
To remedy this,
we are interested in
the stabilization of the index 2 formulation,
possibly along the lines of~\cite{2023KinonBetschSchneider}.
Of course,
the computer implementation
of the EPHS modeling language itself
and
of the presented models within it
are ultimately among our goals.
We also want to mention that
the inspiring work~\cite{2015Sonneville}
also deals with the extension to flexible multibody dynamics
within the $\mathrm{SE}(3)$-based framework.
It can hence be expected that the presented models
can be adapted also to include flexible beams and shells.

\section*{Author contribution statement}%

\textbf{Markus Lohmayer}:
  Conceptualization,
  Investigation,
  Visualization;
  Writing -- Original Draft,
  Writing -- Review \& Editing,
\textbf{Giuseppe Capobianco}:
  Investigation,
  Writing -- Original Draft,
\textbf{Sigrid Leyendecker}:
  Supervision

\section*{Acknowledgements}%

We thank Rodrigo Sato Martin de Almagro
for pointing us to references~%
\cite{1991MarsdenRatiuRaugel,1996BlochKrishnaprasadMarsdenRatiu}.


\bibliographystyle{link-elsarticle-num}
\bibliography{literature}
\addcontentsline{toc}{section}{References}

\end{document}